\documentclass[fleqn,usenatbib,usedcolumn]{mnras}
\usepackage[british]{babel}             % British English hyphenation
\usepackage{newtxtext}                  % Good fonts
\usepackage{newtxmath}                   %   "    "   (non-slanted Greek)
\let\la=\lesssim % for less than similar from newtxmath, not \la from mnras.cls
\let\ga=\gtrsim
\usepackage[T1]{fontenc}                % Good font encoding
\usepackage{bm}
\usepackage[pdftex]{graphicx}
\usepackage{mathtools}
\usepackage{times}
\usepackage{url}
\usepackage{booktabs}
\usepackage{multirow}

\newcommand{\gtc}{\mbox{G$^3$C}}
\newcommand{\hMpc}{\mbox{\,$h^{-1}$\,Mpc}}

\newcommand{\mass}{\mbox{${\cal M}$}}

\newcommand{\volunit}{\mbox{\,$h^{-3}$\,Mpc$^{3}$}}
\newcommand{\denunit}{\mbox{\,$h^{3}$\,Mpc$^{-3}$}}

\newcommand{\lgal}{\textsc{L-Galaxies}}

\title[GAMA: clustering of galaxy groups]
{Galaxy and Mass Assembly (GAMA): the clustering of galaxy groups}

\author[S.D.~Riggs et al.]
{{\parbox{\textwidth}{\raggedright S.D.~Riggs,$^{1}$\thanks{E-mail:~S.Riggs@sussex.ac.uk}
R.~W.~Y.~M.~Barbhuiyan,$^{1}$
J.~Loveday,$^{1}$
S.~Brough,$^{2}$
B.W.~Holwerda,$^{3}$
A.M.~Hopkins,$^{4}$
S.~Phillipps$^{5}$
}}\\
\vspace{0.4cm}\\
{\parbox{\textwidth}{\raggedright 
$^{1}$Astronomy Centre, University of Sussex, Falmer, Brighton BN1 9QH, UK
\\
$^{2}$School of Physics, University of New South Wales, NSW 2052, Australia
\\
$^{3}$Department of Physics and Astronomy, University of Louisville, Louisville, KY 40292, USA
\\
$^{4}$Australian Astronomical Optics, Macquarie University
105 Delhi Rd, North Ryde, NSW 2113, Australia
\\
$^{5}$Astrophysics Group, HH Wills Physics Laboratory,
University of Bristol, Tyndall Avenue, Bristol BS8 1TL, UK
\\
}}}

\date{Accepted XXX. Received YYY; in original form ZZZ}

\pubyear{2021}

\begin{document}
\label{firstpage}
\pagerange{\pageref{firstpage}--\pageref{lastpage}}
\maketitle

\begin{abstract}
  We explore the clustering of galaxy groups in the 
  Galaxy and Mass Assembly (GAMA) survey to investigate the dependence 
  of group bias and profile on separation scale and group mass. 
  Due to the inherent uncertainty in estimating the group selection 
  function, and hence the group auto-correlation function, we instead measure
  the projected galaxy--group cross-correlation function. 
  We find that the group profile has a strong dependence on scale 
  and group mass on scales $r_\bot \la 1 \hMpc$.
  We also find evidence that
  the most massive groups live in extended, overdense, structures. 
  In the first application of marked clustering statistics to groups, 
  we find that group-mass marked clustering 
  peaks on scales comparable to the typical group radius of $r_\bot \approx 0.5 \hMpc$.
  While massive galaxies are associated with massive groups, 
  the marked statistics show no indication of galaxy mass segregation within groups. 
  We show similar results from the IllustrisTNG simulations and the \lgal\ 
  model, although \lgal\ shows an enhanced bias and galaxy mass 
  dependence on small scales.
\end{abstract}

\begin{keywords}
galaxies: groups: general --- galaxies: haloes --- large-scale structure of Universe
\end{keywords}

\section{Introduction}
\label{sec:intro}
In the standard hierarchical model of galaxy formation, galaxies form in 
gravitationally collapsed dark matter (DM) haloes which grow by merging with other haloes
\citep[e.g.][]{Press1974, White1978}.
Consequently, the relative density of observable matter ($\delta_g$, such
as galaxies, galaxy groups, and galaxy clusters) in a given volume of
space is believed to trace the relative density of dark
matter, $\delta_m$, in that same space. 
In the linear bias model, $\delta_g = b \delta_m$, where $b$ is known as the
\emph{bias} parameter, which will in general be a function of the tracer population, 
separation scale, and redshift.
This linear bias has previously been shown to increase with halo mass 
\citep[e.g.][]{Mo1996, Sheth1999, Sheth2001, Seljak2004, Tinker2005}.

A direct way to explore the connection between galaxies and their DM haloes
is with galaxy group catalogues. 
The total mass of individual haloes can be estimated using 
the galaxy motions within them \citep[e.g.][]{Girardi1998, Eke2006, Robotham2011}, 
or by scaling relations based on the luminosity or mass of their constituent galaxies
\citep[e.g.][]{Yang2007, Han2015, Viola2015}.
The galaxy distribution within haloes can be explored directly by group stacking
\citep[e.g.][]{Budzynski2012} 
or with group-galaxy clustering \citep[e.g.][]{Wang2008, Mohammad2016}.
Group clustering probes intermediate scales compared to the 
typical galaxy- and galaxy cluster-scales used in most clustering studies,
and can be combined with galaxy- and dark matter-clustering to extract estimates of bias.

The mass and colour-dependence of the clustering and bias of galaxy groups was
investigated for SDSS Data Release 4 \citep{Adelman-McCarthy2006}
by \citet{Wang2008}.
They found that the clustering strength of groups increases with increasing total group mass
and also that groups of comparable mass are more strongly clustered when
they contain redder galaxies.
Similar results from SDSS were found in the earlier study by \cite{Berlind2006}, 
where a sharp increase in group-galaxy clustering is observed 
within the typical group scale compared to larger scales.
Further, group-galaxy clustering is observed to decrease slightly on scales $r_\bot \la 0.3 \hMpc$,
possibly suggesting the existence of group cores,
although clustering measurements on these scales are sensitive to the choice of group centre,
a point we discuss further in section~\ref{sec:cens}.
An increase in clustering strength with increasing group mass has also been shown
at slightly higher redshifts using the zCOSMOS survey \citep{Lilly2007}
by \cite{Knobel2012}.

While there is a general consensus in previous work on the group
clustering increase with group mass at large scale, the details on small scales are less constrained.
A key aspect to this is the dependence of the positions of galaxies 
within groups on the properties of the satellite galaxies.
Mass segregation, a tendency for more massive galaxies to 
be closer to the group centre, is found by, e.g.,
\citet{Presotto2012, Roberts2015},
but other studies \citep[e.g.][]{VonDerLinden2010, Kafle2016}
find no trend in stellar mass with radial distance from group centre.
The presence or absence of mass segregation helps constrain the strength
of dynamical friction effects within haloes, as satellites infall at large radii
\citep{Wetzel2013} and then move inwards due to dynamical friction.

Standard two-point clustering measurements can be expanded on using 
marked statistics 
\citep{Stoyan1994, Beisbart2000, Sheth2004, Sheth2005, Skibba2006, Harker2006, White2009, White2016}.
These have been used to explore the environmental dependence of clustering, 
with \cite{Skibba2013} finding that small-scale clustering
is dependent on local density,
and \cite{Sheth2004} showing that close pairs of haloes form earlier.
\cite{Armijo2018} show that galaxy clustering has an increasing dependence on
halo mass on smaller scales.
However, this method has not to our knowledge previously been applied 
to the exploration of group clustering.

The Galaxy and Mass Assembly
(GAMA; \citealt{Driver2009,Driver2011,Liske2015}) survey provides an
opportunity to reassess the clustering of galaxy groups.
GAMA has a smaller area than SDSS, but provides spectroscopic redshifts
two magnitudes fainter \citep{Hopkins2013}, and is highly complete, 
even in the high-density environments of galaxy groups.
We thus expect the GAMA group catalogue to be more reliable than group
catalogues constructed from SDSS data, and to allow the exploration of
group clustering on much smaller scales.
The clustering of GAMA galaxies has been shown to increase with 
luminosity and mass by \cite{Farrow2015}.
The dependency of galaxy clustering in GAMA on galaxy properties has been explored with
marked correlation functions by \cite{Gunawardhana2018} and
\cite{Sureshkumar2021}, finding that specific star formation rate best traces interactions, and
stellar mass best traces environment.
Within GAMA groups, \cite{Kafle2016} find negligible mass segregation for satellites.
Recently, \citet[hereafter VM20]{VazquezMata2020} explored the stellar masses and \textit{r}-band
luminosities of galaxies in GAMA groups, finding brighter and more massive galaxies in 
more massive groups.

In this paper, we present group--galaxy cross-correlation functions
from the GAMA survey; exploring their dependence on scale and group mass.
We consider both the large, inter-group, scales which can be compared to results from SDSS,
and the smaller, intra-group, scales that are opened up with the high completeness of GAMA.
We further examine these dependencies by presenting the first application of 
marked correlations to group clustering.
We also compare these correlation functions to results from the IllustrisTNG hydrodynamical simulations
\citep{Marinacci2018,Naiman2018,Nelson2018,Nelson2019,Pillepich2018,Springel2018}
and the \lgal\ semi-analytic model \citep{Henriques2015}.

The layout of this paper is as follows: 
in Section 2 we describe the data selection from the GAMA survey, mock catalogues
and models we compare against; 
in Section 3 we detail the methods used to derive the two-point correlation functions 
and marked statistics;
in Section 4 we present our results;
and finally in Sections 5 and 6 we provide a discussion and conclusion.
The cosmology assumed throughout is that of a $\Lambda$CDM model with
$\Omega_{\Lambda} = 0.75$, $\Omega_{\textrm{m}} = 0.25$,
and $H_{0} = h 100 \textrm{km s}^{-1}\textrm{Mpc}^{-1}$.
We represent group (halo) masses on a logarithmic scale by 
$\lg {\cal M}_{h} \equiv \log_{10}({\cal M}_{h}/{\cal M}_{\odot}h^{-1})$,
where we take ${\cal M}_{h}$ to be $M_{200}$, defined by the 
mass enclosed within an overdensity 200 times the mean density of the universe.

\section{Data,  mocks, and simulations} \label{sec:data}
The GAMA data and mock catalogues used in this analysis are identical to those
used in a recent study of the dependence of the galaxy luminosity and stellar
mass functions on the mass of their host groups (VM20),
although we select groups and galaxies using different mass and redshift cuts.
We summarise the most salient features here.

We make use of the GAMA-II \citep{Liske2015} equatorial fields G09, G12 and G15,
centred on 09h, 12h and 14h30m RA respectively.
These fields each have an area of 12 $\times$ 5 degrees, Petrosian
magnitude limit of $r < 19.8$ mag,
and a completeness greater than 96\% for all
galaxies which have up to 5 neighbours within 40 arcsec; 
for a more in-depth description see \cite{Liske2015}.

\subsection{Galaxy sample}

It is necessary to use a volume-limited sample of galaxies for cross-correlating
with groups, as more massive groups are at higher redshift,
where galaxies in a flux-limited sample will be more luminous and
therefore more strongly clustered.
In other words, using a flux-limited galaxy sample,
apparent clustering strength would increase with halo mass,
even if there was no dependence of halo bias on mass.

We select a volume-limited sample of 42,679 GAMA-II galaxies which have
($K + e$)-corrected $r$-band Petrosian magnitude $^{0.1}M_r < -20$ mag,
with corresponding redshift limit $z_{\rm lim} < 0.267$ and mean number density
$n = 5.38 \times 10^{-3} \denunit$.
This corresponds to the `V0' sample of
\citet[hereafter L18]{Loveday2018}\footnote{
  Attentive readers will notice that here we use a slightly higher redshift
  limit for the same absolute magnitude limit as L18.
  This is due to an alternative way of defining a 95 percent complete sample.
  In L18, we take the 95th-percentile of the $K$-correction of galaxies
  within $z_{\rm lim} \pm 0.01$.
  Here we take the 95th-percentile of the projected $K$-correction
  $K(z_{\rm lim})$ of all galaxies with $z < z_{\rm lim}$.
}, and is chosen to
roughly maximise survey volume and number of galaxies.
We choose to define the volume-limited sample by luminosity
rather than stellar mass, as (i) the parent sample is magnitude limited,
meaning that variations in mass-to-light ratio would require much more
stringent cuts on mass than on luminosity, and (ii) estimated stellar mass
is inherently more uncertain (and model-dependent) than luminosity.

To account for the different redshifts at which galaxies are observed,
the intrinsic luminosities of the GAMA galaxies we use have been corrected by
the so-called $K$-correction \citep{Humason1956}.
We obtain $K$-corrections from the GAMA data management unit (DMU)
{\tt kCorrectionsv05}; see \cite{Loveday2015} for details on how these were calculated.
We $K$-correct to a passband blue-shifted by $z = 0.1$ in order to minimize
the size, and hence uncertainty, in $K$-correction.
Absolute magnitudes in this band are indicated by $^{0.1}M_r$.
We include luminosity evolution by applying correction of $+Q_e z$ mag, where $Q_e = 1.0$.

\begin{table*}
  \caption{Definition of galaxy volume-limited samples for GAMA data,
    mocks, TNG300-1 simulation, and \lgal\ SAM.
    The columns are absolute $r$-band magnitude limit 
    ($K$-corrected to redshift 0.1 for GAMA, redshift 0.0 for other samples), 
    redshift limit, sample volume,
    number of galaxies selected, and mean density.
    GAMA data and mocks cover areas of 180 and 144 deg$^2$ respectively.
    The mock sample was volume-limited to redshift 0.301 before 
    applying the GAMA redshift limit, leading to a slightly higher final number density.
    TNG300-1 and \lgal\ use periodic boxes,
    and so are volume-limited by nature.
    The redshifts we quote for them are those of the output snapshot used.
\label{tab:gal_def}}
\begin{math}
\begin{array}{cccccc}
  \hline
  & M_{\rm lim} & z_{\rm lim} & V  & N_{\rm gal} & \bar{n} \\
  &  &  & [10^6 \volunit] &  & [10^{-3} \denunit]\\
  \hline
  {\rm GAMA} & -20.00 & 0.267 & 7.93 & 42,679 & 5.38 \\
  {\rm Mock} & -20.21 & 0.267 & 6.35 & 34,615 & 5.45 \\
  \mbox{TNG300-1} & -19.83 & 0.200 & 8.62 & 46,349 & 5.38 \\
  {\rm \lgal} & -20.12 & 0.180 & 110.78 & 596,023 & 5.38 \\
  \hline
\end{array}
\end{math}
\end{table*}

The statistics of the GAMA volume-limited galaxy sample,
along with those of the mock catalogue and simulations,
are summarized in Table~\ref{tab:gal_def}.
The GAMA data, mocks and simulations have galaxies with differing
$K$ and $e$-corrections, and different luminosity functions,
but we only need luminosities
in order to generate comparable volume-limited galaxy samples.
We therefore choose magnitude limits (shown in the second column of
Table~\ref{tab:gal_def}),
in order to achieve approximately equal number densities (final column),
and hence clustering properties.
Note that the GAMA mocks were designed to have a luminosity function very close to that of the
GAMA data (R11).  The different magnitude limits in Table~\ref{tab:gal_def}
most likely reflect differences in the $K$- and $e$- corrections assumed, as well as
sample variance in the GAMA data \citep{Driver2011}.
For reference, the clustering and stellar mass distribution
for our GAMA, mock, and simulated galaxy samples are presented in 
Appendix~\ref{appen:lf}.

\subsection{GAMA groups}

The GAMA Galaxy Group Catalogue (\gtc v9) was produced by 
grouping galaxies in the GAMA-II spectroscopic survey using a friends-of-friends (FoF) algorithm; 
this is an updated version of \gtc v1 which was
generated from the GAMA-I survey by \citet[][hereafter R11]{Robotham2011}.
The FoF parameters used for \gtc v9 (hereafter abbreviated to \gtc) 
are identical to those in R11, but applied to the larger GAMA-II galaxy sample.
\gtc\ contains 23,654 groups with 2 or more members and
overall $\sim 40\%$ of galaxies in GAMA are assigned to \gtc\ groups.
In this study, we utilise only those groups within the redshift limit
$z < 0.267$ of our volume-limited galaxy sample (Table~\ref{tab:gal_def}),
and which have five or more member galaxies,
as R11 find these richer groups are most reliable.
Reducing the threshold on the number of group members increases the
number of low-mass groups, but these groups are very unreliable as 
chance alignments are increasingly included in the group sample.
We also require groups in our sample to have {\tt  GroupEdge} $> 0.9$,
this removes any where it is estimated that less than 90\% of the group
is within the GAMA-II survey boundaries.
This leaves us with a sample of 1,894 groups with $12.0 < \lg {\cal M}_{h} < 14.8$.
We do not attempt to form a volume-limited sample of GAMA groups,
as selection effects are complex (see VM20 for a discussion),
and to do so would severely limit the number of groups that could be used.

We take the centre of these groups to be the iterative central from R11,
found by iteratively removing galaxies from the centre of light until one is left.
We choose this as it is found by R11 to be the best estimator of true central,
but we discuss the choice of this further in section \ref{sec:cens}.

Halo masses ${\cal M}_h$ are estimated from group $r$-band luminosity
(column {\tt  LumB}) using the scaling relation for $M_{200}$ of
\citet[][eqn.~37]{Viola2015}, which was calibrated against weak-lensing measurements.
The {\tt  LumB} column contains total $r$-band luminosities down to
$M_r - 5 \log_{10} h = -14$ mag
in solar luminosities, corrected by an empirical factor $B$ which has been
calibrated against mock catalogues (see R11 section 4.4 for details).
The \gtc\ also provides dynamical mass estimates derived via the virial theorem (column {\tt MassA}).

Our choice of luminosity-based mass estimates follows the
checks on mass estimate reliability by VM20,
who find that the luminosity-based estimates correlate much better 
with true halo mass than dynamical mass estimates (VM20 Fig. 1).

\begin{figure*} 
\includegraphics[width=0.9\linewidth]{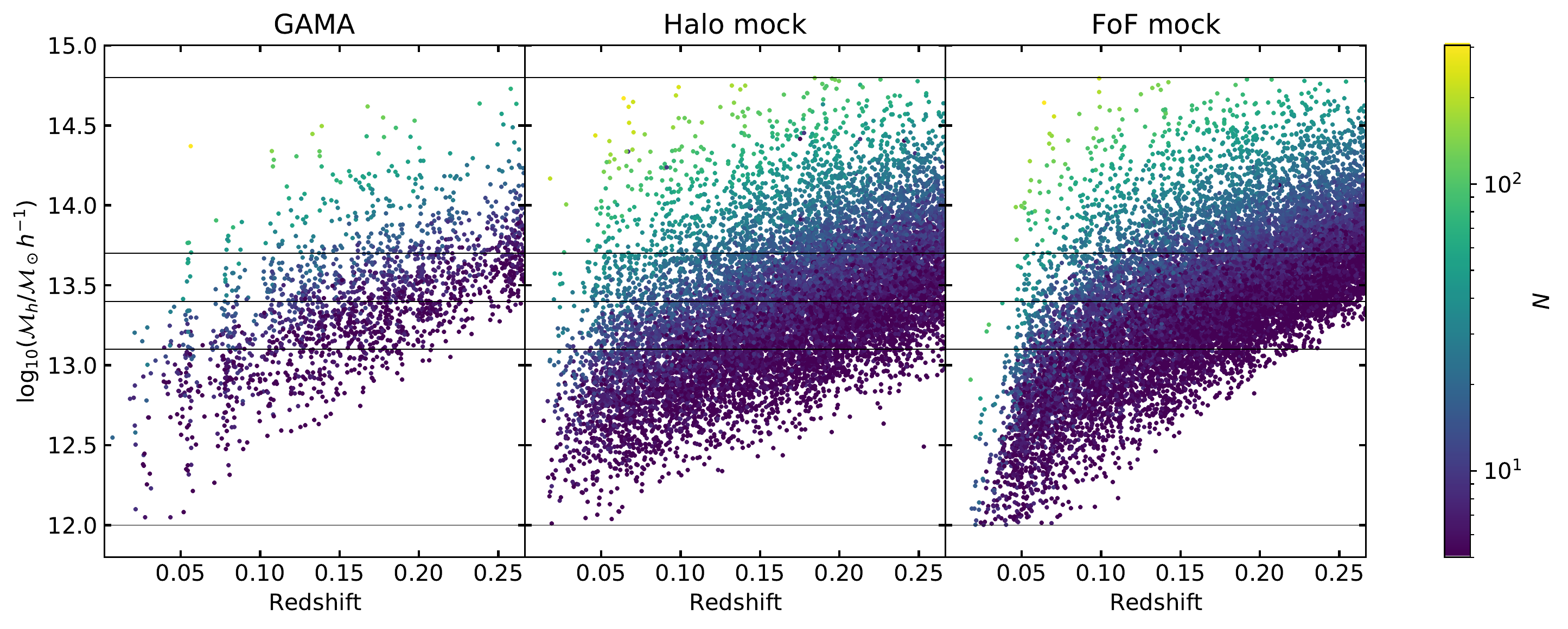} 
\caption{
  Mass--redshift distribution for GAMA and mock groups 
  at $z < 0.267$ with at least 5 members.
  Groups are colour-coded by the number of group members.
  The horizontal lines show the division of groups into halo mass bins.
  Mock groups are shown for all 9 realisations of the lightcone.
  }
\label{fig:mass_z}
\end{figure*}

We show the mass--redshift distribution of our selected GAMA groups in 
the left panel of Fig.~\ref{fig:mass_z}.
Due to the $r < 19.8$ mag flux limit of GAMA-II 
and our requirement for groups to contain at least 5 members,
low-mass groups are less likely to be detected at higher redshifts,
and the groups that are detected generally have fewer observed members.

\begin{table*}
\caption{Group bin names and log-mass limits, number of groups,
mean log-mass, and mean redshift for GAMA-II data, intrinsic mock haloes,
FoF mock groups, TNG300-1 simulation haloes and \lgal\ SAM haloes.
Note that the values given for mocks are averages across the 9 realisations used.
TNG300-1 and \lgal\ are results from single snapshots 
down-sampled to select groups comparable to GAMA.
\label{tab:group_mass_def}}
\begin{tabular}{cccccccccccccccccccc}
\hline
& & \multicolumn{3}{c}{GAMA} & & \multicolumn{3}{c}{Halo Mocks} & &
\multicolumn{3}{c}{FoF Mocks} & & \multicolumn{2}{c}{TNG300-1} & & \multicolumn{2}{c}{\lgal}\\
\cline{3-5} \cline{7-9} \cline{11-13} \cline{15-16} \cline{18-19}\\[-2ex]
& $\lg {\cal M}_{h, {\rm limits}}$ & $N$ & $\overline{\lg {\cal M}}$ &
$\overline z$ & &
$N$ & $\overline{\lg {\cal M}}$ &  $\overline z$ & &
$N$ & $\overline{\lg {\cal M}}$ &  $\overline z$ & &
$N$ & $\overline{\lg {\cal M}}$ & &
$N$ & $\overline{\lg {\cal M}}$ \\
\hline
${\cal M}1$ & [12.0, 13.1] & 380 & 12.87 & 0.10 & & 352 & 12.86 & 0.11 & & 346 & 12.80 & 0.10 & & 414 & 12.84 & & 5276 & 12.84\\
${\cal M}2$ & [13.1, 13.4] & 547 & 13.26 & 0.15 & & 383 & 13.25 & 0.16 & & 401 & 13.26 & 0.15 & & 383 & 13.25 & & 4986 & 13.25\\
${\cal M}3$ & [13.4, 13.7] & 566 & 13.54 & 0.19 & & 366 & 13.54 & 0.19 & & 523 & 13.55 & 0.19 & & 405 & 13.54 & & 5127 & 13.55\\
${\cal M}4$ & [13.7, 14.8] & 401 & 13.93 & 0.20 & & 306 & 13.97 & 0.20 & & 430 & 13.96 & 0.21 & & 461 & 13.98 & & 6263 & 14.00\\
\hline
Total & [12.0, 14.8] & 1894 & 13.41 & 0.16 &  & 1407 & 13.39 & 0.17 &  & 1699 & 13.43 & 0.17 &  & 1663 & 13.42 &  & 21652 & 13.44 \\
\hline
\end{tabular}
\end{table*}

We sub-divide the groups into four mass bins as defined in
Table~\ref{tab:group_mass_def},
chosen as a compromise between bins of fixed mass range and comparable
group numbers.
As seen in VM20, the central galaxy luminosity is greater for more massive groups,
with our mass bins \mass1--4 having central galaxy mean absolute magnitudes of
$^{0.1}M_r - 5 \log_{10} h = -20.48$, $-21.12$, $-21.48$, and $-21.87$, respectively.
We note that this means that the \mass1 centrals have a lower mean luminosity
than our volume limited galaxy sample, which has a mean $^{0.1}M_r - 5 \log_{10} h = -20.59$,
and so the \mass1 groups are expected to be slightly less clustered than the galaxy sample.

\subsection{Mock catalogues}

\begin{figure} 
\includegraphics[width=\linewidth]{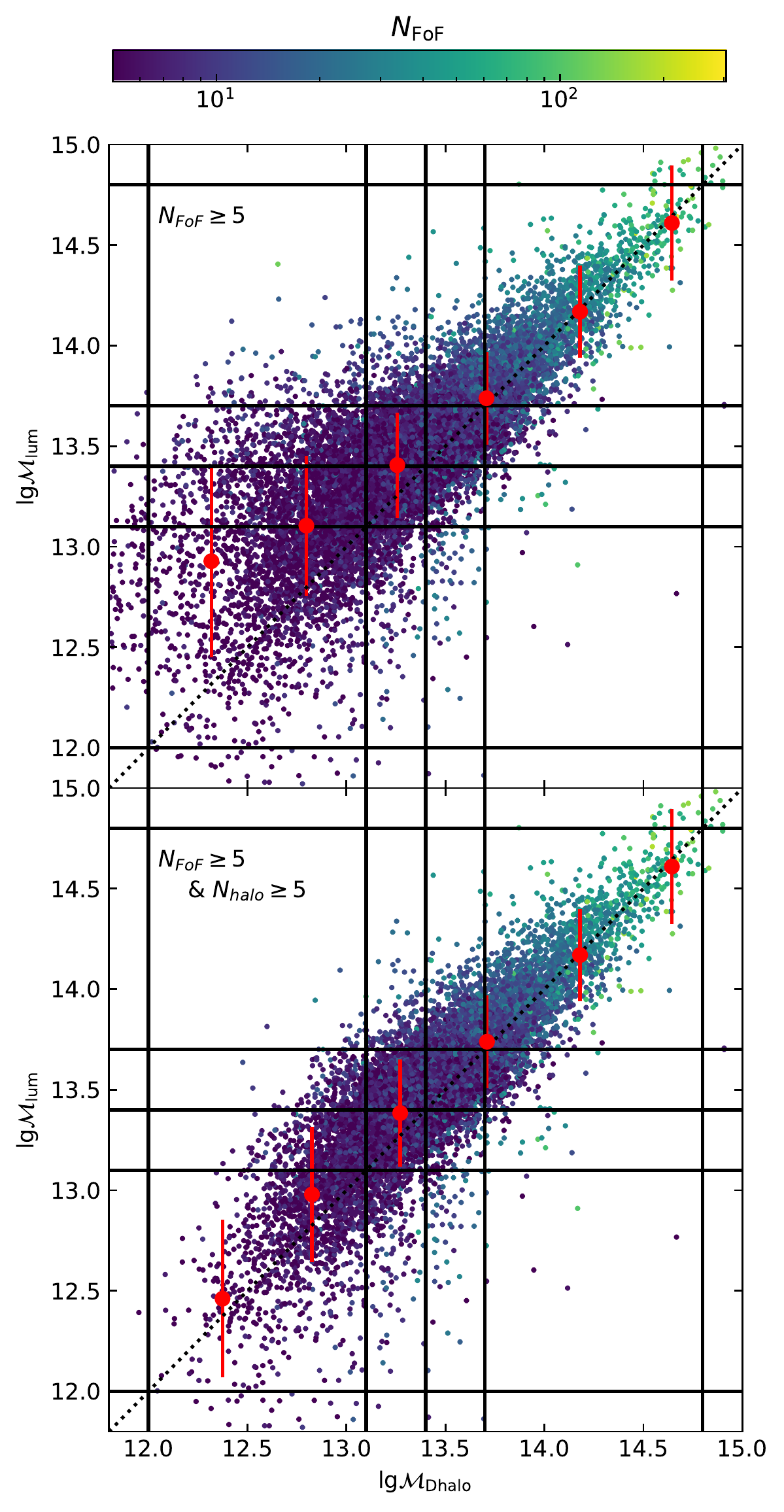} 
\caption{Comparison of luminosity-based ($\lg {\cal M}_{\rm lum}$) estimates of
  mock group mass, against true mock halo mass ($\lg {\cal M}_{\rm Dhalo}$),
  colour coded by group membership,
  for groups at redshifts $z<0.267$.
  The upper panels show groups selected by their visibility in the FoF mocks
  ($N_{\rm FoF} \geqslant 5$), while the lower panels show only those groups
  visible in both mocks ($N_{\rm FoF} \geqslant 5$ and $N_{\rm halo} \geqslant 5$).
  The red error-bars show mean and standard deviation of $\lg {\cal M}_{\rm lum}$
  in 0.5 mag bins of $\lg {\cal M}_{\rm Dhalo}$.
  The horizontal and vertical lines delineate the halo mass bins used in this analysis.
  }
\label{fig:mass_comp}
\end{figure}

We compare our results with predictions from two sets of mock group catalogues 
for the GAMA-I survey (catalogues updated to the GAMA-II survey are currently being developed).
These catalogues were produced using lightcones from the {\sc galform} \citep{Bower2006}
semi-analytic galaxy formation model run on the Millennium dark matter simulation
\citep{Springel2005}.
For more details on these mocks we refer the reader to R11.

The first set of mocks are {\tt G3CMockHaloGroupv06}, which we refer to as \emph{halo mocks}.
This contains the dark matter haloes in the simulations,
with their positions and masses $\mass_{\rm Dhalo}$.
The definition of $\mass_{\rm Dhalo}$ differs slightly from $M_{200}$,
but \cite{Jiang2014} and R11 find they are median unbiased relative to each other,
so we can use $\mass_{\rm Dhalo}$ as an estimate of $M_{200}$.
The second set of mocks are {\tt G3CMockFoFGroupv06}, which we refer to as \emph{FoF mocks}.
The groups in this are generated with the same FoF algorithm as GAMA,
and masses $\mass_{\rm lum}$ estimated using the same \citet{Viola2015}
luminosity scaling relation.
Comparing results from these two mock group catalogues thus allows us to
assess the impact on estimated halo clustering of redshift-space group-finding
and luminosity-based mass estimation.
For halo and FoF mock groups that share a common central galaxy,
Fig.~\ref{fig:mass_comp} compares $\mass_{\rm lum}$ with $\mass_{\rm Dhalo}$.
The upper panel shows all groups with $N_{\rm FoF} \geqslant 5$ 
(and implicitly $N_{\rm halo} \geqslant 2$ to be counted as a group),
while the lower panel shows groups with $N_{\rm FoF} \geqslant 5$ and $N_{\rm halo} \geqslant 5$.
From the lower panel it is apparent there is reasonable agreement of 
$\mass_{\rm lum}$ to $\mass_{\rm Dhalo}$
(within one standard deviation) for groups that have sufficient members 
to be included in our halo catalogue sample.
However, it is clear from the upper panel that there is a population of groups
which have their membership, and therefore mass, overestimated in the FoF mocks.
Through the rest of this work, for consistency with our GAMA selection,
we use all groups in the FoF mock with $N_{\rm FoF} \geqslant 5$,
so the groups with overestimated mass are included.
In the halo mock we select all groups with $N_{\rm halo} \geqslant 5$,
representing the sample we would have if the FoF group finder 
perfectly assigned galaxies to groups.

The central and right panels of Fig.~\ref{fig:mass_z}, showing the mass-redshift
relation for all selected mock groups, further shows that the luminosity-based masses
have a stronger redshift dependence than the true halo masses.
The mass overestimation appears to be greater at high redshift.
However, at redshifts $z \la 0.04$ the FoF mock groups mostly have low masses,
suggesting galaxies are missed from the outskirts of the more extended groups 
at low redshift.
We expect this to imply a similar trend in group mass misestimation with redshift 
will also be present in the groups from GAMA.

Mock galaxies are taken from the galaxy catalogue 
associated with the mock groups we use, {\tt G3CMockGalv06}.
We $K$-correct the absolute magnitudes to redshift zero with
the $K$- and $e$-corrections specified in Sec.~2.2 of R11.
Due to differences with GAMA $K$- and $e$-corrections, we set the galaxy
magnitude limit by trial-and-error to give approximately the same mean volume-limited
number density as the GAMA galaxy sample.
This results in a sample with a limiting absolute magnitude $^{0.0}M_r < -20.21$
and limiting redshift $z_{\rm lim} < 0.301$. 
The typical masses of observed galaxies and groups increase with redshift, 
and so to ensure that the mock samples are comparable to the observations,
we then restrict our mock sample to the GAMA redshift limit of $z_{\rm lim} < 0.267$. 
The details of our final mock galaxy sample are given in Table~\ref{tab:gal_def}.

We estimate uncertainties on mock clustering from the scatter
between nine realisations of the GAMA-I survey equatorial regions.
Each of these realisations consists of three $12 \times 4$ deg regions;
which are 20 per cent smaller in area (and so also volume) than the equatorial fields we use from GAMA-II.
Galaxy stellar masses are not included in these mocks so we
cannot explore the dependence of the marked correlation on galaxy mass in the mocks.

\subsection{Random catalogues} \label{sec:randoms}

A random sample of points is needed to model any selection effects in the
galaxy sample (our choice of cross-correlation estimator in section \ref{sec:xcorr}
means that the selection function of group samples is not needed).
We use the same survey mask described in section 2.3.1 of L18,
and generate angular coordinates using {\sc mangle}
\citep{Hamilton2004,Swanson2008}.
Radial coordinates are drawn at random from a uniform distribution in 
comoving volume with a modulation factor of $10^{0.4 P z}$, the density-evolution factor of
\citet[][equation 5]{Loveday2015}, taking $P = 1$.
We generate 10 times more random points than galaxies.

\begin{figure} 
\includegraphics[width=\linewidth]{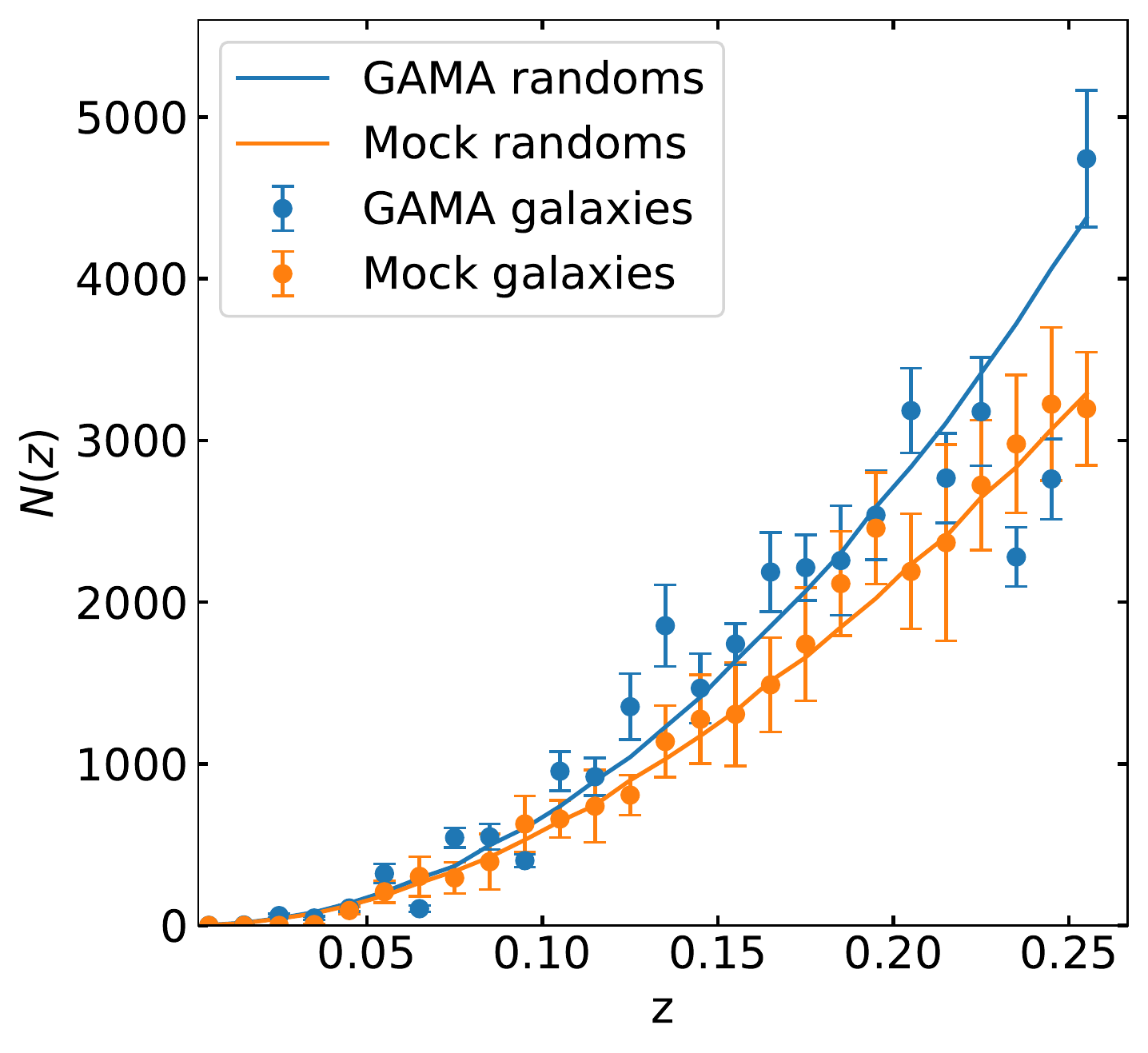} 
\caption{Comparison of galaxy redshift distributions for GAMA, 
    the average across the nine mocks, and the random samples.
    Random counts have been divided by 10 to account for the 
    larger number of random points generated.
    Uncertainties on GAMA and random counts are found by jackknife between 27 regions in RA,
    and on the mock counts by the scatter between 9 realisations. 
    The offset in the number of galaxies between GAMA and the mocks is 
    due to the larger area of GAMA (180 degrees$^2$ compared to 144 degrees$^2$).
  }
\label{fig:galaxies_Nz}
\end{figure}

In Fig.~\ref{fig:galaxies_Nz} we show galaxy redshift distributions
for GAMA, the average across the nine mocks, and random samples
(with the number of randoms divided by 10 to match the data samples).
The random number counts accurately reproduce the GAMA redshift distribution
except for fluctuations due to large-scale structure 
\citep[c.f.][Fig.~7]{Loveday2015}.

\subsection{Comparison models}
\label{sec:sims}

In addition to comparison with GAMA mock catalogues, we also compare our results
with predictions from the IllustrisTNG hydrodynamical simulations
\citep{Marinacci2018,Naiman2018,Nelson2018,Nelson2019,Pillepich2018,Springel2018}
and the \cite{Henriques2015} version of the \lgal\ semi-analytic model.
For each of these, we select galaxies at  a snapshot close to the GAMA mean redshift,
selecting $z=0.20$ in IllustrisTNG and $z=0.18$ (the closest snapshot to $z=0.20$) in \lgal,
and set the absolute magnitude limit 
of the galaxy sample in order to give the same approximate number density 
as the GAMA volume-limited sample, viz.,
$n = 5.38 \times 10^{-3} \denunit$.

For IllustrisTNG, we use the highest resolution simulation at the largest box-size
of 300 Mpc ($205 \hMpc$ for $h = 0.6774$), TNG300-1.
Haloes are selected by $M_{200}$ (\verb|Group_M_Mean200|)
using the mass limits in Table~\ref{tab:group_mass_def}.
For galaxy masses we select the stellar component (type 4) of the 
{\verb SubhaloMassInRadType } field, which gives the stellar mass within
twice the stellar half mass radius.
Following the recommendation of \cite{Pillepich2018},
we multiply these by a factor of 1.4, 
appropriate for haloes in the mass range $12 < \mathrm{lg}{\cal M}_h < 15$.
We use the dust-corrected luminosities derived from dust model C of \cite{Nelson2018}
when selecting the volume-limited galaxy sample.

For \lgal, we use the \cite{Henriques2015} version with
the Millennium \citep{Springel2005} N-body simulation.
Haloes are again selected by $M_{200}$ and the total stellar mass
of the galaxies is taken.

To avoid including galaxies below the resolution limits of the TNG300-1 and \lgal\ simulations, 
we select only galaxies with $\log_{10} \mathcal{M}_{\star} > 9.0 \mathcal{M}_{\odot}$.

To provide comparable group samples, we need to allow for the fact that
the periodic-cube (i.e. volume-limited) simulations contain many more low-mass groups than
the flux-limited GAMA data and mocks.
We describe here our approach to the group selection; 
in Appendix~\ref{appen:selection} we validate our method 
and demonstrate the consequences of not applying it.

Since we are measuring only group-galaxy cross-correlation functions,
we do not require the simulated groups to have an accurate group auto-correlation.
Therefore, rather than attempt to create lightcones from the simulations,
we simply down-sample the simulated groups to match the mass distribution of selected GAMA groups.
We do this by estimating the probability of 
finding each halo\footnote{The terms `halo' and `group' are used interchangeably
when discussing the TNG and \lgal\ simulations.} within the GAMA volume.
In our GAMA sample we have set $N_{\rm FoF} \geqslant 5$, and so the 
halo selection probability is dependent on
the fifth brightest galaxy in the halo.
To calculate this probability and select simulated groups we use the following procedure for each halo:
\begin{enumerate}
    \item Identify the absolute magnitude of the fifth brightest galaxy in the halo.
    \item Calculate the luminosity distance (and corresponding comoving distance) 
    at which this galaxy would have an observed magnitude of $m_r = 19.8$ mag, the GAMA limit.
    \item Calculate the volume of the GAMA lightcone out to this comoving distance.
    \item Divide by the total volume of our GAMA sample to get the selection probability.
    \item Multiply selection probabilities by 0.95 to account for the use of
        a $95\%$ complete sample based on \textit{K}-corrections 
        in GAMA (we do not attempt to model \textit{K}-corrections for simulated galaxies).
    \item Assign a random number to the halo and include the halo in our sample 
        if this is less than the selection probability.
\end{enumerate}

\subsection{Comparison of group samples}

\begin{figure} 
\includegraphics[width=\linewidth]{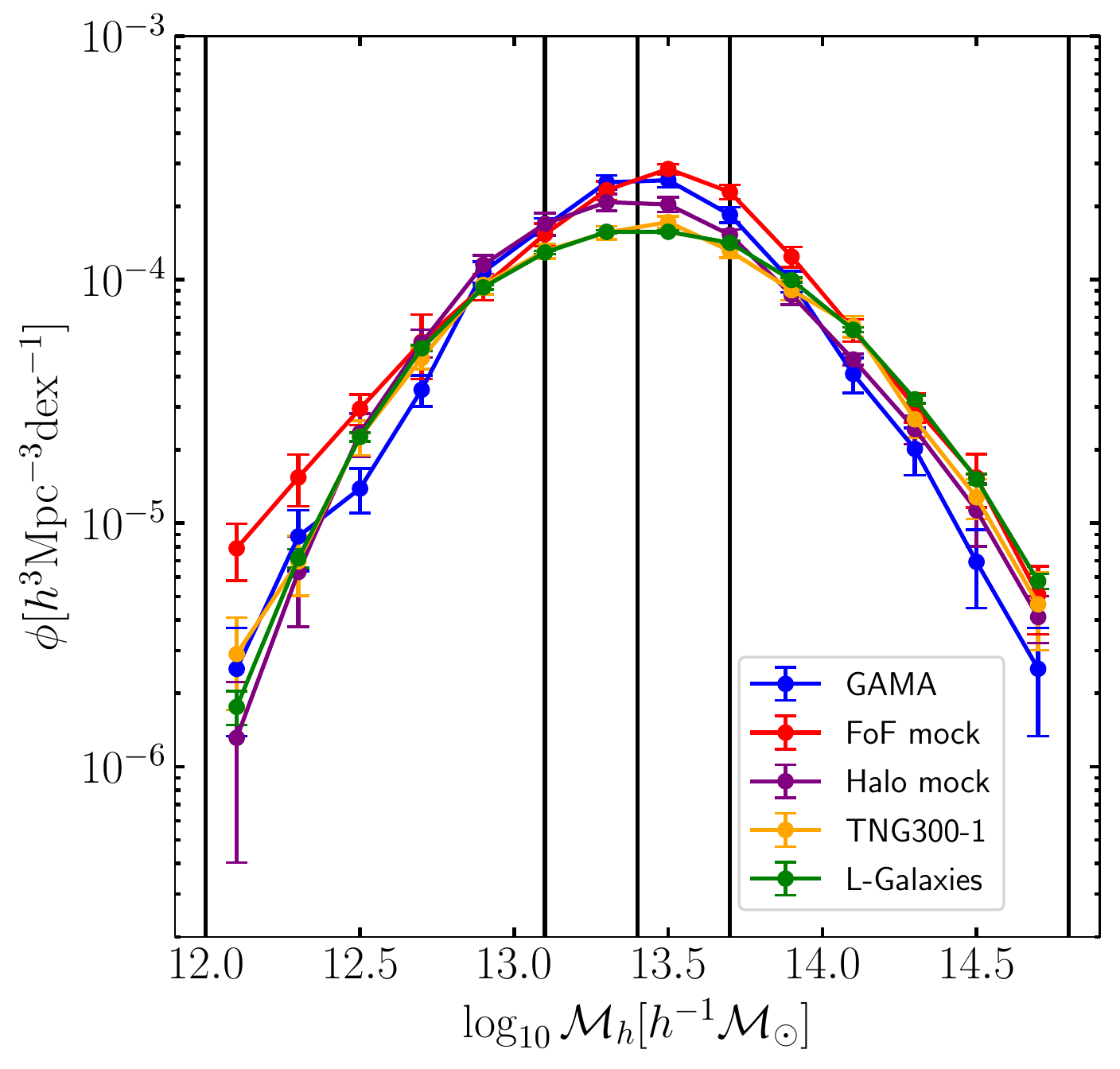} 
\caption{Distribution of group (halo) masses in our sample for GAMA, the two mock catalogues, 
    TNG300-1 and \lgal.
    The plotted uncertainties are jackknife values between 27 regions for GAMA and simulations,
    and the scatter between 9 realisations for the mocks.
    The vertical lines delineate the halo mass bins used in this analysis.
    }
\label{fig:hmf}
\end{figure}

Statistics for the groups selected in GAMA, the mocks and the comparison models
are tabulated in Table~\ref{tab:group_mass_def}.
To complement this, the group mass distributions are shown in Fig.~\ref{fig:hmf}.

The GAMA group masses display a strongly peaked distribution,
with more groups in \mass2 and \mass3 than the other bins.
Comparing the halo and FoF mock groups, it is clear from the table that the FoF algorithm
is systematically overestimating the numbers of groups for the two higher mass bins.
The slightly lower mean mass of \mass1 FoF versus halo groups is likely due to the fact
that $\mass_{\rm lum}$ is systematically underestimated for low redshifts 
where these low-mass haloes are found (see Fig.~\ref{fig:mass_z}).
For higher-mass haloes, $\mass_{\rm lum}$ correlates well with $\mass_{\rm Dhalo}$,
(see Fig.~\ref{fig:mass_comp}),
and so it seems likely that the higher numbers of larger-mass FoF groups
is due to the FoF algorithm aggregating lower-mass haloes into one system.

Comparing the FoF mock groups with the GAMA groups,
it is clear that the mock groups in the lowest mass bin tend to be of slightly
lower mass than the corresponding GAMA groups,
and of higher mass than the GAMA groups in the highest mass bin.
It also appears that relatively there are slightly more high- than low-mass 
groups in the FoF mocks.
These differences should be borne in mind when comparing results from GAMA
data and mock catalogues.

TNG matches the halo mock well on both the mean group masses 
and the mass distribution of selected haloes.
The only apparent difference is in the relative numbers of groups in bin \mass4
compared to the other bins, with TNG showing a greater relative number.
This demonstrates the success of our group selection in TNG, 
which has the predominant effect of removing low mass groups.

\lgal\ matches the halo mock mean group masses and 
follows very similar trends to TNG.
It can be seen from Fig.~\ref{fig:hmf} that the 
mass distribution is almost identical to that from TNG,
except for a slightly greater number of haloes 
at the highest mass end.

\section{Measuring the correlation function} \label{sec:xi2d}
We estimate the galaxy auto-correlation function and group--galaxy
cross-correlation functions in bins of halo mass, as well as
marked correlation functions, in which we weight groups and/or
galaxies by their estimated mass.

We use {\sc Corrfunc} \citep{Sinha2019,Sinha2020} to calculate pair counts
for the clustering statistics.
When plotting correlation functions, we always plot $w_{p}$ against the
{\em mean separation} of galaxy pairs in each bin,
rather than the centre of each (log-spaced) bin.

\subsection{GAMA data and mock catalogues}
\label{sec:xcorr}

In order to overcome the effects of redshift space distortions in the lightcones,
we start by estimating the two-dimensional 
group--galaxy cross--correlation function $\xi_{Gg}(r_\bot, r_\|)$
and galaxy auto--correlation function $\xi_{gg}(r_\bot, r_\|)$;
the excess probability above random of finding 
a group and a galaxy (cross--correlation)
or two galaxies (auto--correlation)
separated by $r_\|$ along the line of sight (LOS) 
and $r_\bot$ perpendicular to the LOS.
These separations are calculated using the standard method \citep[e.g.][]{Fisher1994}
for pairs of objects with position vectors ${\bm r}_1$ and ${\bm r}_2$.
The separation is given by vector ${\bm s} = {\bm r}_2 - {\bm r}_1$
and the vector to the midpoint of the pair from an observer at the origin by 
${\bm l} = ({\bm r}_1 + {\bm r}_2)/2$.
The separations in the LOS and perpendicular directions are then given by
$r_\| = |{\bm s}.\hat{{\bm l}}|$, with $\hat{{\bm l}}$ being
the unit vector in the direction of ${\bm l}$, and
$r_\bot = \sqrt{{\bm s}.{\bm s} - r_\|^2}$.

Raw pair counts are obtained using {\sc Corrfunc}, then normalised to 
account for the relative total numbers of groups, $N_G$, 
galaxies, $N_g$, and random points, $N_r$.
The normalised galaxy--galaxy, $gg$, group--galaxy, $Gg$, 
group--random, $Gr$, galaxy--random, $gr$,
and random-random, $rr$, pair counts are then used to calculate the
correlation functions.
Specifically, these are obtained by dividing the raw pair counts 
in each separation bin by $N_g^2$, $N_G N_g$, $N_G N_r$, $N_g N_r$, and $N_r^2$,
respectively.

The pair counts may additionally be weighted by 
group and/or galaxy mass in order to obtain marked
correlation functions, and hence explore the dependence of clustering 
on group and galaxy mass.
The random points, which follow the selection function of the galaxy sample,
are generated as described in Section~\ref{sec:randoms}.
A total of 426,790 random points are generated, 
10 times the number of galaxies in the sample.

The galaxy auto--correlation $\xi_{gg}(r_\bot, r_\|)$ is estimated using the standard
\citet{Landy1993} estimator,
\begin{equation}
  \label{eq:ls_Gg}
  \xi_{gg}(r_\bot, r_\|) = \frac{gg - 2gr + rr}{rr},
\end{equation}
while $\xi_{Gg}(r_\bot, r_\|)$ is estimated with the cross-correlation form
\citep{Mohammad2016} of this estimator,
\begin{equation}
  \label{eq:ls_gg}
  \xi_{Gg}(r_\bot, r_\|) = \frac{Gg - Gr - gr + rr}{rr}.
\end{equation}

\begin{figure}
  \begin{center}
    \includegraphics[width=0.9\linewidth]{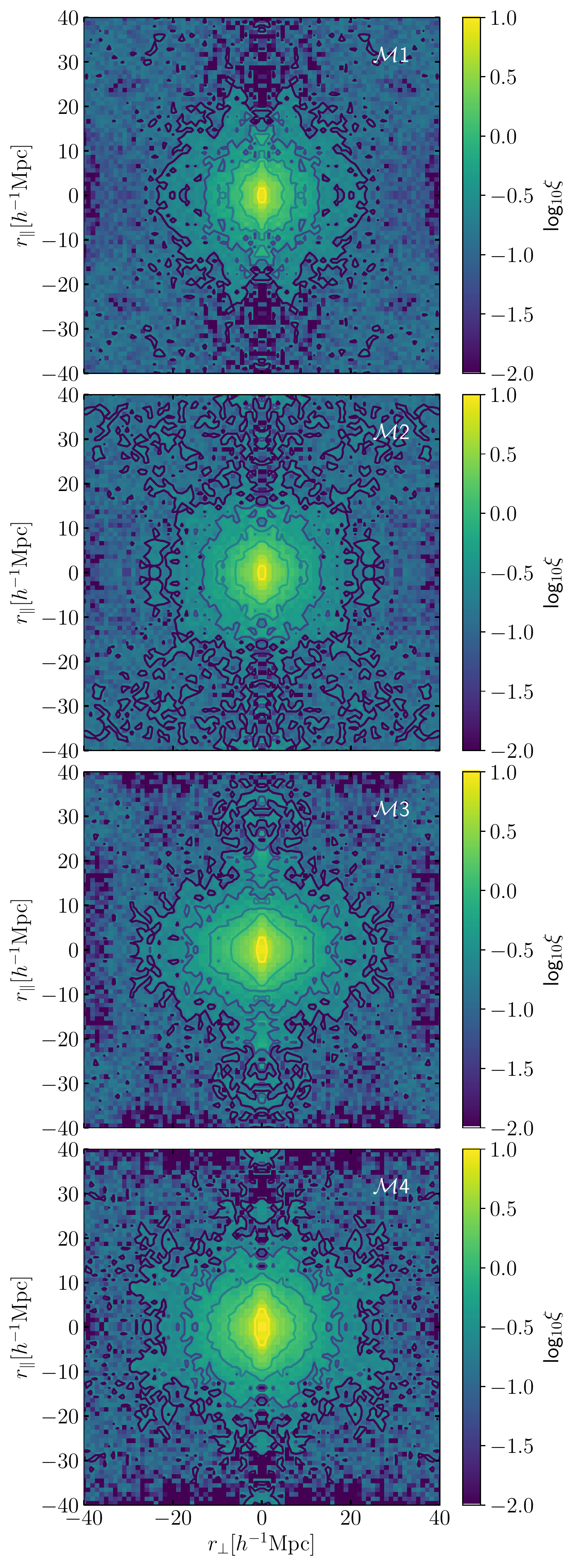}
  \end{center}
  \caption{The two-dimensional group--galaxy cross-correlation functions
  $\xi(r_\bot, r_\|)$
  for our four bins of group mass.
  We show the clustering signal reflected about both axes
  to make it easier to see the distortions introduced by the peculiar velocities
  of galaxies around groups.
  Contour levels are the same as \citet{Li2006}, going up
  from $\xi = 0.1875$ to $\xi = 48$ in factors of 2.
}
\label{fig:xi2d}
\end{figure}

The two-dimensional group--galaxy cross-correlation functions for
our four mass bins of GAMA groups with our volume-limited sample of galaxies,
are shown in Fig.~\ref{fig:xi2d}.
At small projected separations, $r_\bot \la 5 \hMpc$,
the clustering is seen to be stretched along the LOS direction ($r_\|$-axis).
This is increasingly apparent in higher mass bins.
At larger projected separations, the LOS
clustering signal is compressed.

The projected auto- and cross-correlation functions, 
$w_p(r_\bot)$, are obtained by integrating the observed
two-dimensional correlation function $\xi(r_\bot, r_\|)$
along the LOS direction $r_\|$:
\begin{equation}
  \label{eq:proj}
  w_p(r_\bot) = 2 \int_0^{{r_\|}_{\rm max}} \xi(r_\bot, r_\|) d  r_\|.
\end{equation}
We use a limit of ${r_\|}_{\rm max} = 40 \hMpc$;
following the results of \citet[Appendix~B]{Loveday2018}.

To estimate uncertainties on the clustering results from GAMA we use jackknife sampling.
We use 27 regions in RA and calculate error bars as the square root of the 
diagonal terms in the covariance matrix calculated from these regions.
For the mock catalogues 9 different realisations are available and 
we estimate uncertainties using the scatter between these.

The jackknife sampling we use is designed to reproduce the cosmic variance
between independent regions. This accurately reproduces the uncertainty on large scales,
and on small scales can be interpreted as an upper bound on the variation between groups.

\subsection{Simulations}

TNG and \lgal\ use periodic boxes with no redshift space distortions, 
and so we can directly calculate the three-dimensional correlation function $\xi(r)$
using the simplified formula
\begin{equation}
\xi (r) = \frac{DD}{RR} - 1, 
\end{equation}
with the normalised data pair count DD 
(Gg for the cross--correlation, gg for the auto--correlation) 
and random pair count RR.
We again make use of {\sc Corrfunc} to calculate the data pair counts,
normalised by total galaxy and group numbers as above.

Due to periodic boundary conditions, no random catalogue is needed.
Instead, the normalised random pair count is calculated as 
\begin{equation}
RR = \frac{v(r)}{V} ,
\label{eq:sim_RR}
\end{equation}
where \it V \rm is the total box volume and 
$v(r) = \frac{4}{3} \pi ( (r+dr)^3 - r^3)$ is the volume of a 
spherical shell of radius \it r  \rm and thickness \it dr \rm  \citep{Alonso2012}.

The real-space three-dimensional correlation function $\xi(r)$
is then converted to a projected correlation function using
\begin{multline}
w_p (r_\bot) = 2 \int_0^{y_{\rm max}} \xi \left( (r_\bot^2 + y^2)^{1/2} \right) dy \\
= 2 \int_{r_\bot}^{r_{\rm max}} \frac{r \xi(r)}{\sqrt{r^2 - r_\bot^2}} dr,
\label{eq:sim_w}
\end{multline}
to produce a quantity directly comparable to the GAMA measurements.
We perform this integral over an interpolation of the $\xi(r)$ and
we again use an upper integration limit of $r_{\rm max} = 40 \hMpc$.
It is pointed out in \cite{VanDenBosch2013} that this integral 
may be biased on large scales relative to clustering calculated from observations, 
but we do not attempt to correct
for this as we are mostly interested in small scales.

To calculate uncertainties in the results for the simulation boxes
we perform jackknife sampling by dividing 
the box into 27 subboxes and excluding these one at a time.
We then give error bars as the square root of the diagonal elements of the covariance matrix.
Jackknife sampling breaks the periodicity of the box, 
and should therefore require a random catalogue.
However, we continue to use equation \ref{eq:sim_RR} for random pair counts,
and account for the changed random-random term by scaling the $\xi(r)$ value in each jackknife region by 
the ratio of the overall $\xi(r)$ in the box against the mean $\xi(r)$ from the jackknife regions.

\subsection{Marked correlation}

The marked correlation $M_w$ is calculated from the unweighted projected two-point correlation function $w_p$ 
and weighted projected two-point correlation function $W_p$ in all cases using \citep{Sheth2005,Skibba2006}
\begin{equation}
    M_w(r_\bot) = \frac{r_\bot + W_p(r_\bot)}{r_\bot + w_p(r_\bot)}.
\end{equation}

Uncertainties on marked correlations would be overestimated if we simply combine the errors on $W_p$ and $w_p$
(see \citealt{Skibba2006}).
Therefore we calculate the marked correlation for each of our jackknife samples separately 
and estimate the uncertainty from these.

\subsection{Bias}

We make use of two bias measures in our analysis.
The first is the relative bias of the group sample compared to the galaxy sample,
which we define as
\begin{equation}
    b_{\rm rel}(r_\bot) = \frac{w_p^{\mbox{Gg}}(r_\bot)}{w_p^{\mbox{gg}}(r_\bot)}.
    \label{eq:rel_bias}
\end{equation}
This accounts for different galaxy auto-correlation amplitudes between samples,
although it does retain some dependence on the galaxy sample.

The second bias measure we use is that relative to dark matter.
We define the galaxy bias $b_{g}$ using
\begin{equation}
    w_p^{\mbox{gg}} (r_\bot) = b_{g}^2 (r_\bot) w_p^{\rm DM} (r_\bot),
    \label{eq:abs_gal_bias}
\end{equation}
and the corresponding group bias $b_{G}$ with
\begin{equation}
    w_p^{\mbox{Gg}} (r_\bot) = b_{G} (r_\bot) b_{g} (r_\bot) w_p^{\rm DM} (r_\bot).
    \label{eq:abs_grp_bias}
\end{equation}
Note that in this notation the relative bias from equation \ref{eq:rel_bias} becomes
$b_{\rm rel} = b_{G} / b_{g}$.

For the dark matter auto-correlation, $w_p^{\rm DM}$, we use the Millennium simulations, 
the Millennium \citep{Springel2005} $\xi(r)$ on scales $r > 1 \hMpc$ and 
Millennium-II \citep{BoylanKolchin2009} on smaller scales,
which we project from $\xi(r)$ to $w_p(r_\bot)$ 
by interpolating $\xi(r)$ and using equation \ref{eq:sim_w},
the same method we used for the simulation correlations.

\section{Results}
\label{sec:results}
\subsection{FoF versus halo mocks}

\begin{figure}
\includegraphics[width=\linewidth]{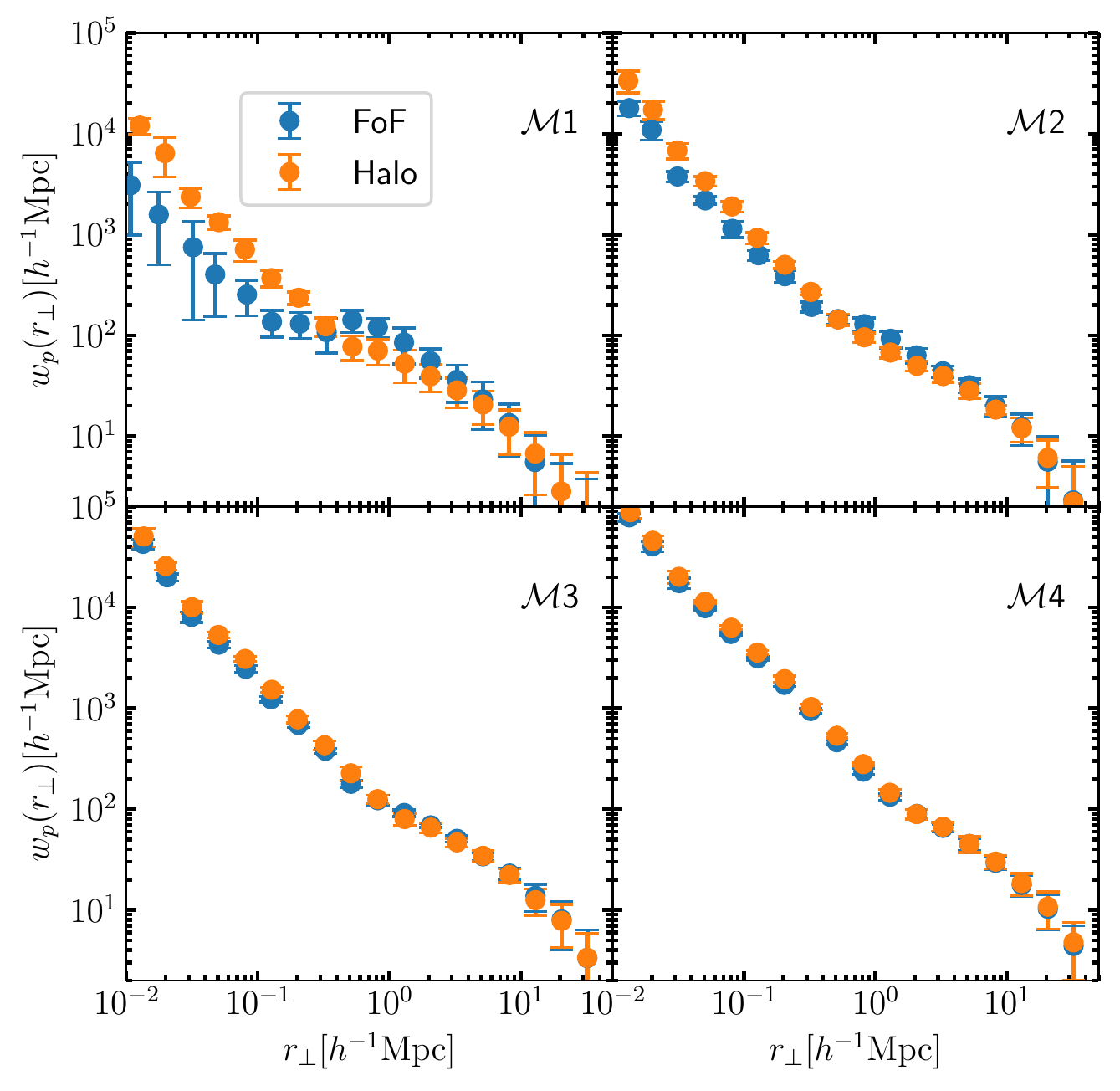}
\caption{Group-galaxy cross-correlation functions for the mock catalogues.
  Orange symbols show results using the halo mocks,
  blue symbols show results obtained using FoF mocks.
}
\label{fig:wp_mock}
\end{figure}

We first compare clustering results obtained using the FoF and halo mocks
in Fig.~\ref{fig:wp_mock}.
We see that in mass bins 3 and 4, the FoF mock group clustering is in
very good agreement with that of the halo mocks, despite the large excess
of FoF groups in these mass bins (Table~\ref{tab:group_mass_def}).
However, for the lower mass bins, particularly \mass1,
the FoF group clustering is
underestimated on very small scales, $r_\bot \la 0.2 \hMpc$,
and very slightly overestimated on scales
$0.5 \la r_\bot \la 2 \hMpc$.
It seems likely that the low mass FoF groups may be contaminated by
chance projections of isolated galaxies,
thus reducing the small-scale clustering signal.
Insofar as the mock catalogues are representative of the GAMA data,
we can infer that the GAMA results are likely to be reliable in mass bins 2--4,
but that those for \mass1 should be treated with some scepticism.

\begin{figure}
\includegraphics[width=\linewidth]{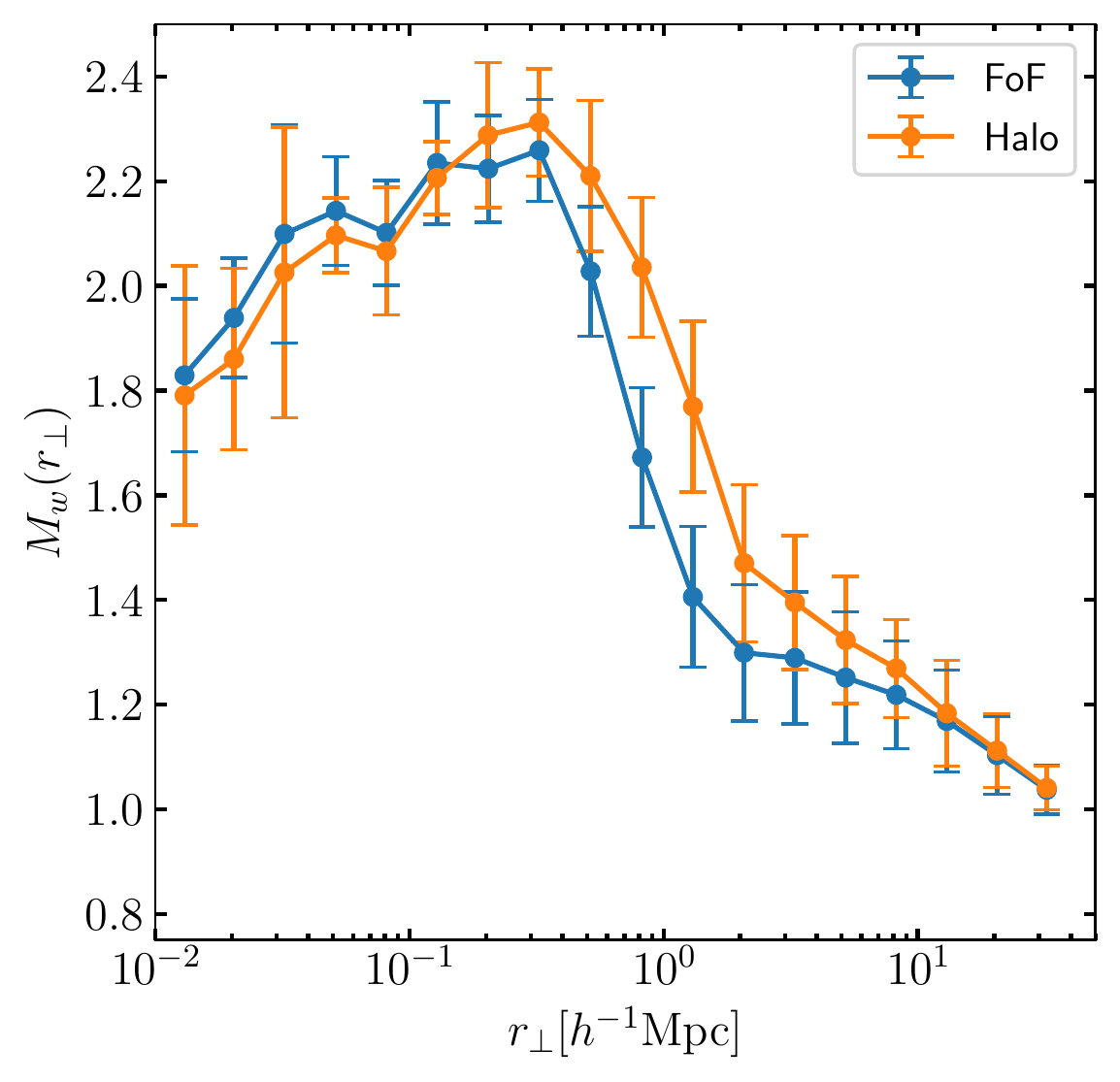}
\caption{Marked correlation for the mock catalogues,
  using group mass as the mark.
  Orange symbols show results using the halo mocks,
  blue symbols show results obtained using FoF mocks.
}
\label{fig:marks_mock}
\end{figure}

To check the effects of the group finding on the marked correlation,
we show the group mass marked correlation for the FoF and halo mocks in 
Fig.~\ref{fig:marks_mock}.
We see that on small scales the mocks agree, but on scales
$r_\bot \ga 0.1 \hMpc$ the FoF marked correlation is lower.
This is around the size of a compact group, and is perhaps due both
to spurious FoF groups (created by chance alignments) being isolated from other galaxies,
and also to more extended groups being missed by the FoF group finder.
We expect this trend to be representative of GAMA, 
and so the GAMA marked correlation may also be biased low on these scales.

\subsection{Group clustering and bias in mass bins}

\subsubsection{GAMA and mocks}

\begin{figure*}
\includegraphics[width=\linewidth]{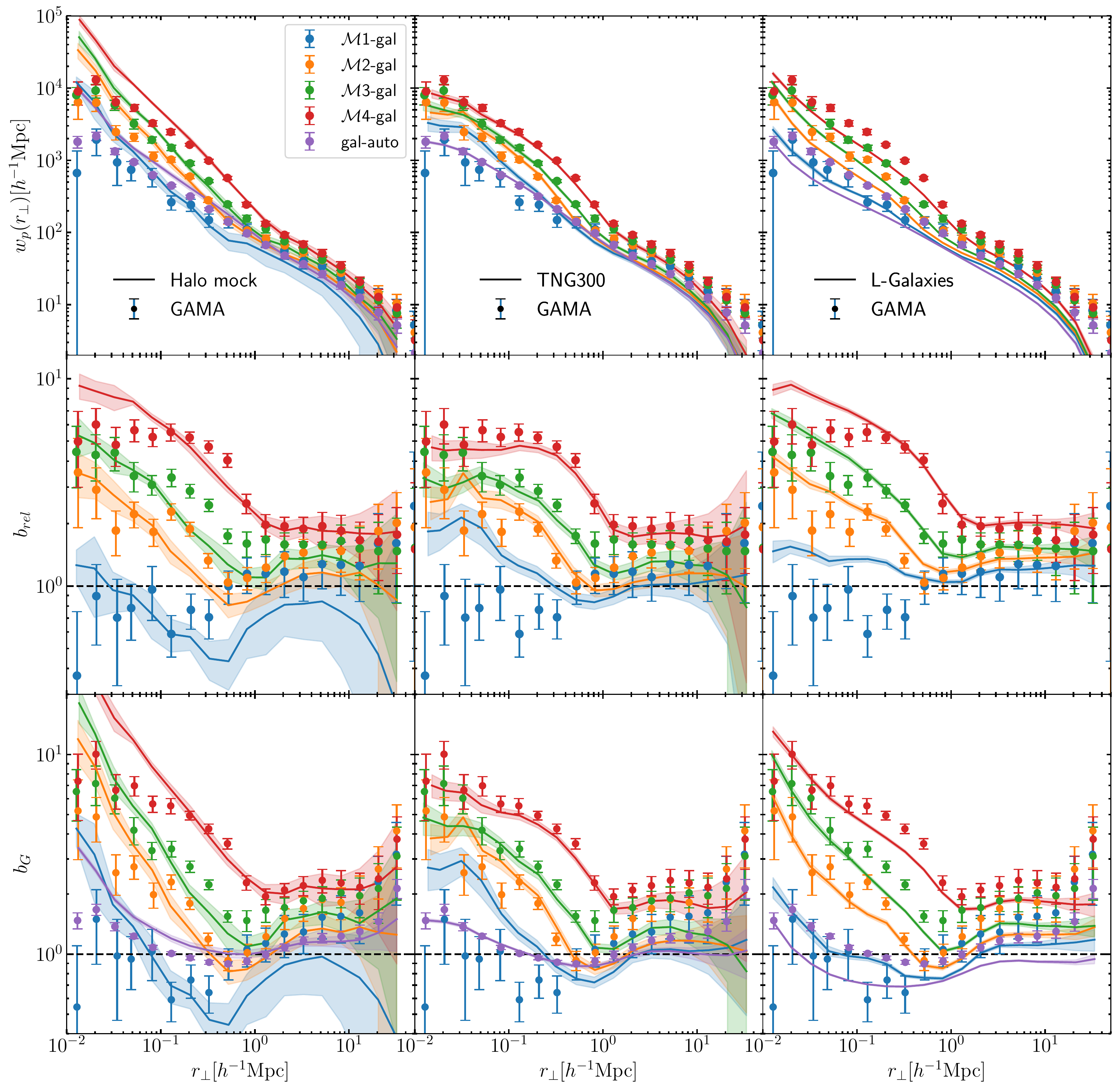}
\caption{Top panels: The projected group--galaxy cross-correlation functions
  for our four bins of group mass as indicated.
  Also shown is the galaxy auto-correlation function.
  Middle panels: Relative bias of the projected group--galaxy cross-correlation to the 
  galaxy sample, obtained by dividing the group--galaxy cross-correlation by the
  galaxy auto-correlation.
  Bottom panels: Bias of the projected group--galaxy cross-correlation 
  and galaxy auto-correlation relative to the
  dark matter auto-correlation function of the Millennium simulations.
  In all panels, symbols and error bars show the GAMA results;
  lines of corresponding colour show results from the halo mock in the left panels,
  the Illustris TNG300-1 simulation in the central panels, and \lgal\ in the right panels.
}
\label{fig:w_p_bias}
\end{figure*}

Fig.~\ref{fig:w_p_bias} shows the GAMA projected group--galaxy cross-correlation
functions for each group mass bin (top), 
along with the bias relative to the galaxy sample (middle),
and the bias relative to a DM-only simulation (bottom).
Left, middle and right panels show comparison results from the halo mocks,
TNG, and \lgal, respectively.
Both bias estimates are highly dependent on scale and group mass
on  intra-group scales, $r_\bot \la 1 \hMpc$.
On larger scales, the biases are relatively constant
(within the error bars) for each mass bin, but there is still a slight trend
for bias to increase with mass.

On scales $r_\bot \approx 0.1 \hMpc$, GAMA relative group bias 
($b_{\rm rel}$ from equation \ref{eq:rel_bias}; middle panels)
increases rapidly with group mass,
from $b \approx 0.8 \pm 0.2$ for \mass1 groups to $b \approx 5 \pm 1$
for \mass4 groups.
The strong halo-mass dependence of small scale clustering seen here is to be expected,
as on scales $r_\bot \la 1 \hMpc$, the cross-correlation signal will be dominated
by galaxies within each respective halo (intra-halo clustering) and
group membership increases with halo mass.

Comparison in Fig.~\ref{fig:wp_mock} of the FoF and halo mocks on these scales suggests that the
apparent below-unity bias of \mass1 groups is partly an artefact of the group-finding
algorithm, although it also reflects a lack of bright galaxies in these small groups. 
The halo mock bias in the middle-left panel of Fig.~\ref{fig:w_p_bias} 
is consistent with unity for \mass1 groups at $r_\bot \approx 0.1 \hMpc$,
although it drops below unity above this, reaching a minimum at $r_\bot \approx 0.5 \hMpc$,
indicating the spatial extent of these smaller groups.
As with other mass bins, the mock galaxies in \mass1 groups seem to be too centrally-concentrated.

On larger scales (1--5 $\hMpc$), the dependence of relative bias on group mass
is weaker, although the bias of the highest mass bin is still 2--3
times that of the lowest mass groups.
By scales of $r_\bot \approx 10 \hMpc$, the biases of each mass bin are
consistent within the uncertainties.

On the largest scales $r_\bot \ga 10 \hMpc$, the relative bias remains constant in each
bin within uncertainties but the GAMA auto- and cross-correlation functions are seen to have 
slightly greater amplitude than those of the mocks.
This perhaps indicates small differences in the galaxy populations used,
but as these scales are also the most affected by the projection of the clustering
signal, we cannot draw any firm conclusions on these scales.

When turning to bias relative to the dark matter auto-correlation 
($b_{g}$ and $b_{G}$ from equation \ref{eq:abs_grp_bias}; lower panels), 
the bias is seen to increase down to the smallest scales we 
plot for the galaxies and the groups in bins \mass2--4.
As with the bias relative to the galaxies, \mass1 GAMA groups show a bias
of about unity on the smallest scales not seen in the halo mock, 
which is likely to be a result of the group-finding algorithm.

The halo mocks substantially over-predict the bias on small scales.
On intra-halo scales the relative bias (middle panels) is seen to increase 
roughly as a power-law with
decreasing $r_{\bot}$, rather than displaying a flattening as seen in GAMA.
This becomes even more apparent in bias relative to the dark matter (lower panels),
with an even steeper increase to small scales.
This suggests inaccuracy in the physics defining satellite galaxy occupation
and positions in the mocks, with satellites being placed too close to the centre on average. 
This is perhaps unsurprising given the uncertainties in the modelling of satellite
mergers when the dark matter subhalo they are associated with disappears \citep[see e.g.][]{Pujol2017}.

\subsubsection{TNG300 and \lgal}

In Fig.~\ref{fig:w_p_bias}, we also show corresponding results from the Illustris TNG300-1
simulation and the \lgal\ semi-analytic model, each around the mean GAMA redshift $z=0.2$.

TNG results are shown as solid lines in the central column of panels.
The TNG galaxy auto-correlation function (purple line)
is in very close agreement with GAMA on scales $r_\bot \la 5 \hMpc$, 
although slightly below that of the mock,
and the TNG halo--galaxy cross-correlation functions show a similar 
characteristic inflection to GAMA
around $r_\bot \approx 0.5$--$1 \hMpc$; the transition 
from the intra-halo to the inter-halo regime.
In the higher mass bins, \mass3 and \mass4, the amplitude of the cross-correlations
is also in agreement with GAMA on smaller scales within uncertainties.
In \mass1, and to a lesser extent in \mass2, for which GAMA results are suspect,
TNG shows a greater cross-correlation on scales $r_\bot \la 0.3 \hMpc$ than GAMA.
This is clearest moving to the smallest scales, $r_\bot \la 0.05 \hMpc$, where it leads
to convergence of \mass1--\mass3 results as \mass1 and \mass2 continue to rise
while \mass3 and \mass4 flatten off.

Solid lines in the right-hand panels of Fig.~\ref{fig:w_p_bias} show results for \lgal.
Both the galaxy auto-correlation and halo--galaxy cross-correlations fall below
the GAMA results.
The relative biases in \lgal\ show the trend seen in the halo mock of a 
continuing increase down to the smallest scales and greater
amplitude than GAMA, suggesting the same issues in the two SAMs.
However, the group bias in the lower panels agrees well 
with GAMA on scales $r_\bot \ga 0.1 \hMpc$, 
implying some of the discrepancy is connected to the galaxy sample.
This difference in the dependence on the galaxy properties between \lgal\ 
and GAMA becomes clearer in the marked correlations discussed below.
On larger scales, \lgal\ shows the halo mass dependence of bias continuing beyond $r_\bot = 5 \hMpc$,
showing the most massive groups are at the centre of denser regions extending further 
than those of smaller groups, in agreement with GAMA.

\subsection{Marked correlation functions}

\begin{figure*}
\includegraphics[width=\linewidth]{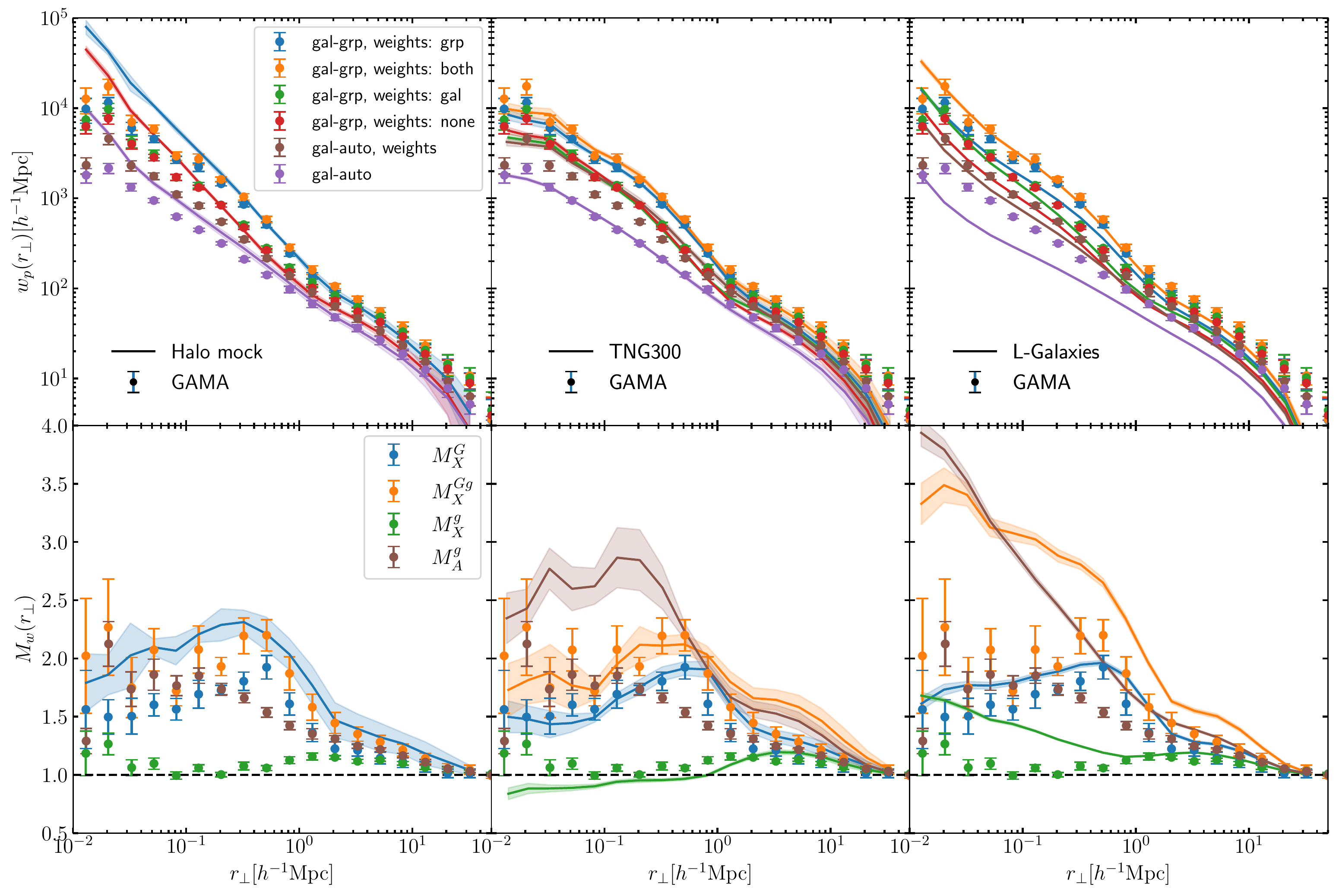}
\caption{
  Top panels: The projected group--galaxy cross-correlation functions
  for all groups, weighted by galaxy and group masses as indicated.
  Also shown is the galaxy auto-correlation function both unweighted 
  and using galaxy masses as weights.
  Bottom panels: Marked cross-correlations using galaxy masses ($M_X^g$), 
  group masses ($M_X^G$), and both masses ($M_X^{Gg}$) as marks,
  along with the stellar-mass marked galaxy auto-correlation ($M_A^g$).
  In all panels, symbols and error bars show the GAMA results;
  lines of corresponding colour show results from the halo mock in the left panels,
  the Illustris TNG300-1 simulation in the central panels and \lgal\ in the right panels.
}
\label{fig:marked_corr}
\end{figure*}

\subsubsection{Marked cross-correlation}

The upper panels of Fig.~\ref{fig:marked_corr} show projected correlation functions 
weighted in the various ways indicated.
Lower panels show marked group--galaxy cross-correlation functions 
using group mass ($M_X^G$), galaxy mass ($M_X^g$), and both masses ($M_X^{Gg}$) as weights,
as well as the marked galaxy auto-correlation function ($M_A^g$).
We weight by linear mass in order to enhance the differences
between the marked statistics, although the use of log-mass weights does
not qualitatively change our results 
(see \citealt{Sheth2005} for a discussion on re-scaling marks).
In appendix \ref{appen:ranks} we show, using rank-ordered marks, that the 
specific values of the weights do not affect our conclusions.

The GAMA group-mass marked cross-correlation function ($M_X^G$, blue symbols) peaks at scales 
$r_\bot \approx 0.5 \hMpc$, declining gradually to smaller scales, 
and somewhat more rapidly on larger scales until $r_\bot \approx 2 \hMpc$, 
beyond which $M_X^G$ declines more gradually.
The halo mock (blue line) shows similar trends to GAMA data, 
but with $M_X^G$ about 20 percent higher.
The peak in $M_X^G$ around $r_\bot \approx 0.5 \hMpc$ is indicative of the 
typical projected radii of our galaxy groups.
It is also consistent with the bias results of Fig.~\ref{fig:w_p_bias},
where the relative strengths of the group biases differ most around this scale,
due to the below-unity bias of \mass1 groups and large bias of \mass4 groups.

The GAMA galaxy-mass marked cross-correlation function
($M_X^g$, green points) is systematically greater than unity
only on inter-group scales, $r_\bot \ga 0.5 \hMpc$.
We are unable to measure $M_X^g$ for the GAMA mocks, as galaxy masses are not available.
When both galaxy and group masses are used as weights ($M_X^{Gg}$, orange points), 
a slight additional enhancement is seen relative to $M_X^G$,
indicative of the most massive groups having an enhanced number of massive satellite galaxies.

$M_X^G$ measurements from both the TNG and \lgal\ simulations show general agreement with GAMA.
TNG agrees with GAMA within uncertainties on almost all scales, but is below the mocks
on scales $r_\bot \la 1 \hMpc$. 
\lgal\ on the other hand agrees well on all scales with the halo mock, 
and is generally above but just consistent with the GAMA results. 
The very close agreement between \lgal\ and the halo mock may be a result of both being built upon
the Millennium simulation.

When marking with galaxy masses, TNG shows $M_X^g < 1$ on scales $r_\bot \la 0.5 \hMpc$,
meaning the most massive satellite galaxies are not found near the group centres.
Yet when both group and galaxy masses are used ($M_X^{Gg}$),
an enhancement relative to $M_X^G$ is seen on all scales.
This is consistent with the conclusion from GAMA that the most massive groups also
contain the most massive satellites, but this dependency extends out slightly further in TNG, 
to $r_\bot \approx 10 \hMpc$.

\lgal\ shows a galaxy-mass marked cross-correlation $M_X^g$ greater than unity,
especially on scales  $r_\bot \la 1 \hMpc$ where $M_X^g$ is seen 
to increase as scale decreases, meaning massive satellites are always closely associated 
with the group centre.
The same trend is seen and enhanced even further when both group and galaxy masses are 
used as marks ($M_X^{Gg}$).
This is consistent with the high small-scale bias we observed for \lgal, 
yet very different from the GAMA result,
suggesting that the satellite galaxies in \lgal\ are typically more massive.
This is in accord with the finding in VM20 that the modified Schechter functions appropriate for
GAMA satellite galaxies under-predict the number of massive satellites in \lgal.

\subsubsection{Marked auto-correlation}

For GAMA, \lgal\ and TNG, we also show the (stellar mass) marked galaxy auto-correlation
($M_A^g$, brown symbols or lines), which helps 
in understanding some of the differences in the group-galaxy cross-correlations.
GAMA shows no systematic scale-dependence (but large scatter)
in $M_A^g$ on scales $r_\bot \la 0.2 \hMpc$, 
but then declines systematically on larger scales, 
always lying below $M_X^{Gg}$.
This makes sense, as $M_A^g$ indirectly contains group information through
the presence of central galaxies, although these will have lower masses than the groups.

TNG on the other hand shows a marked auto-correlation $M_A^g$ which peaks on scales
0.1--0.5 \hMpc\ and decreases slightly on smaller scales.
The large enhancement compared to GAMA and the TNG cross-correlation functions is 
likely to be due to the apparent over-dependence
of central galaxy mass on group mass in TNG reported by VM20.
The decreasing dependence on the smallest scales is consistent with the trends 
in $M_X^g$, and shows that the most massive galaxies have a slight tendency to avoid group centres.

\lgal\ shows a very different trend that the most massive galaxies are very close together, 
with $M_A^g$ still increasing at $r_\bot \approx 0.01 \hMpc$.
This matches the cross-correlation result and also appears consistent with a slight trend 
in \cite{Henriques2017} for the auto-correlation to be below SDSS in lower mass bins and 
above in higher mass bins.
This is likely to be the result of the supernova feedback used, as \cite{VanDaalen2016} find
that the feedback strength affects the relative proportions of satellite galaxies of 
different masses.

The general picture found from the marked correlations is one of agreement in the group mass dependence
of clustering, but disagreement in the galaxy mass dependence.
While the group mass dependence is a significant success in the positioning of galaxies 
within groups in both TNG and \lgal,
massive galaxies appear to be too clustered, especially in \lgal.

\section{Discussion}
To put our results into context we discuss here 
the choice of group centre, which is the main caveat to our work,
and compare against previous works.

\subsection{Choice of group centre}
\label{sec:cens}

\begin{figure*} 
\includegraphics[width=\linewidth]{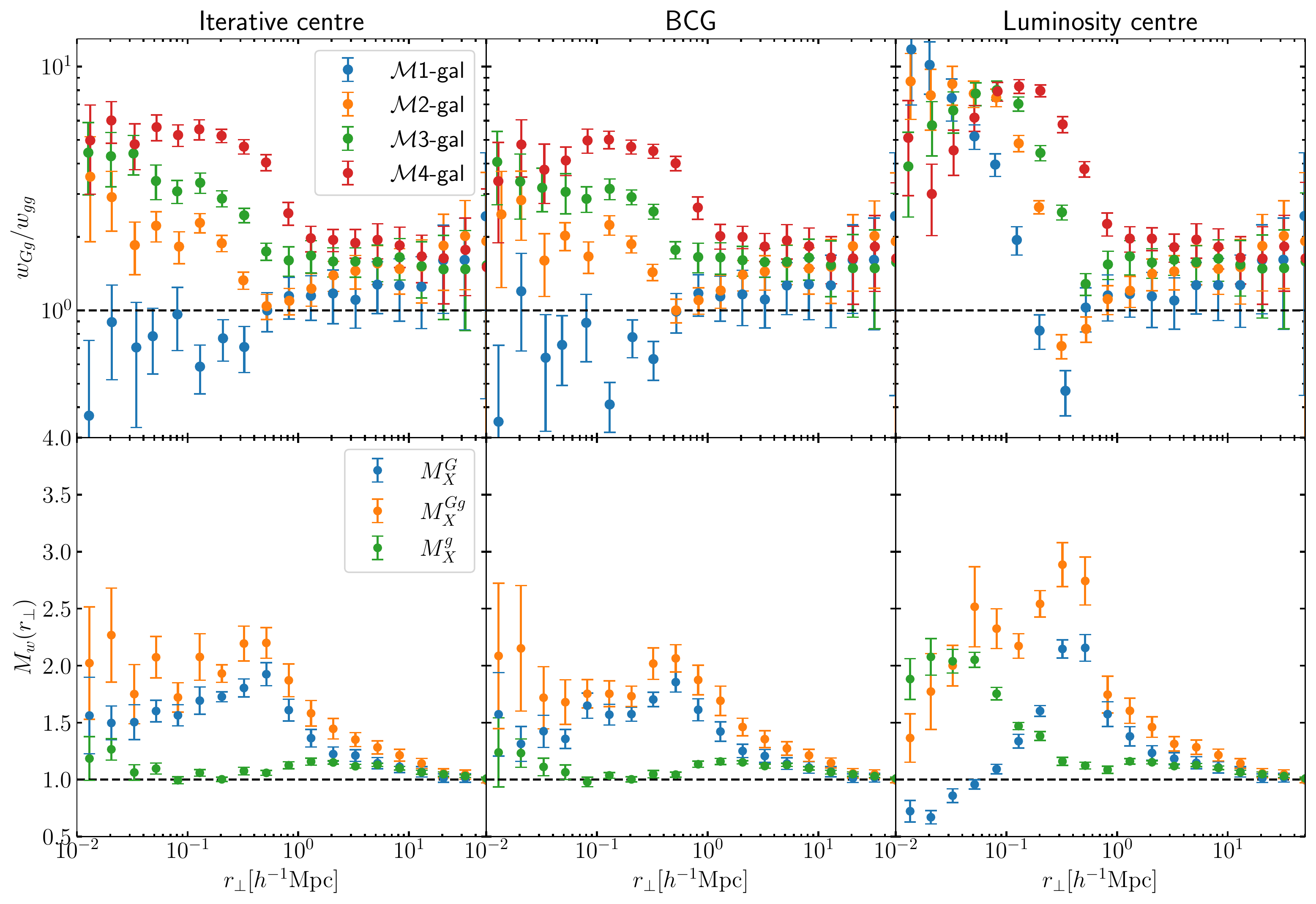} 
\caption{
  Effect of choice of GAMA group centre on our results.
  Upper panels show the relative bias $b_{\rm rel}$ of the projected group--galaxy cross-correlation 
  to the galaxy sample, and lower panels show the marked correlation
  using galaxy masses ($M_X^g$), group masses ($M_X^G$), and both masses ($M_X^{Gg}$) as marks.
  The left panels shows the iterative group centre, the middle panels the
  brightest central galaxy, and the right panels the centre-of-light of the group.
  }
\label{fig:cen_bias}
\end{figure*}

In this work we have considered group--galaxy cross-correlation functions
in GAMA down to scales smaller than the typical group size, 
so our results depend heavily on the choice of group centre.
We check here for effects due to possible mis-identification of group centre by using
the three different definitions of group centre described in Section 4.2.1 of R11.

R11 found the most reliable group centre to be the one we have used throughout this work,
the iterative centre.
This was found by iteratively removing the galaxy furthest from the centre-of-light
of all remaining galaxies in the group, until only one galaxy remains.
The position of the final galaxy is taken to be the group centre.
In most cases this is the same as the second definition of group centre, the 
brightest central galaxy (BCG), taken to be the brightest galaxy in the group.
The third definition of group centre corresponds simply to the group centre-of-light, 
which does not in general coincide with a galaxy.
Using mock catalogues, R11 showed the iterative centre to match the true centre in
$\sim 90\%$ of cases, while the BCG showed large offsets in some cases,
and the centre-of-light only matched the true centre for groups where all members are detected.

To explore the effect of group centre choice on our results, we show
in Fig.~\ref{fig:cen_bias} the relative bias $b_{\rm rel}$ of the four group mass bins 
and the marked cross-correlations for the three definitions of group centre.
On the left we show the iterative centre used elsewhere in this work.
This is in most cases the same as the BCG shown in the
middle panel, so the results are similar from these two options.
However, the iterative centre shows a more consistent picture for different group masses
on small scales, while the BCG shows a drop in bias for the most massive groups,
suggesting the galaxy at the centre of the gravitational potential
of the group has been included in the cross-correlation.
The definition of group centre as centre-of-light is shown in the right panel, 
and this definition shows significant evidence of mis-centering.
The bias is seen to be peaked, with the peak at $r_\bot \approx 0.1 \hMpc$ for the most
massive groups, and on smaller scales for less massive groups.
The location of this peak is indicative of the mean offset of the central galaxy
from the centre-of-light.

A similar outcome is found by considering the marked correlations.
Using group mass as the mark, the iterative centre and BCG results are similar,
but the centre-of-light definition shows a negative mark on small scales
related to the reduction in bias for the more massive groups.
When using galaxy masses as marks, the iterative centre and BCG results both show no mark
on scales less than the typical group size, but the centre-of-light shows a positive mark,
probably indicating the inclusion of the the true central galaxy in the cross-correlation.

Based on this, we are in agreement with the result of R11 that the iterative centre 
we have used is the best reflection of the true group centre,
as it does not display the offset in peak bias associated with including the central
galaxy in the cross-correlation.

\subsection{Comparison with previous results}

Finally, we compare our results to previous works,
and calculate the average bias on large scales.

\begin{figure}
\includegraphics[width=\linewidth]{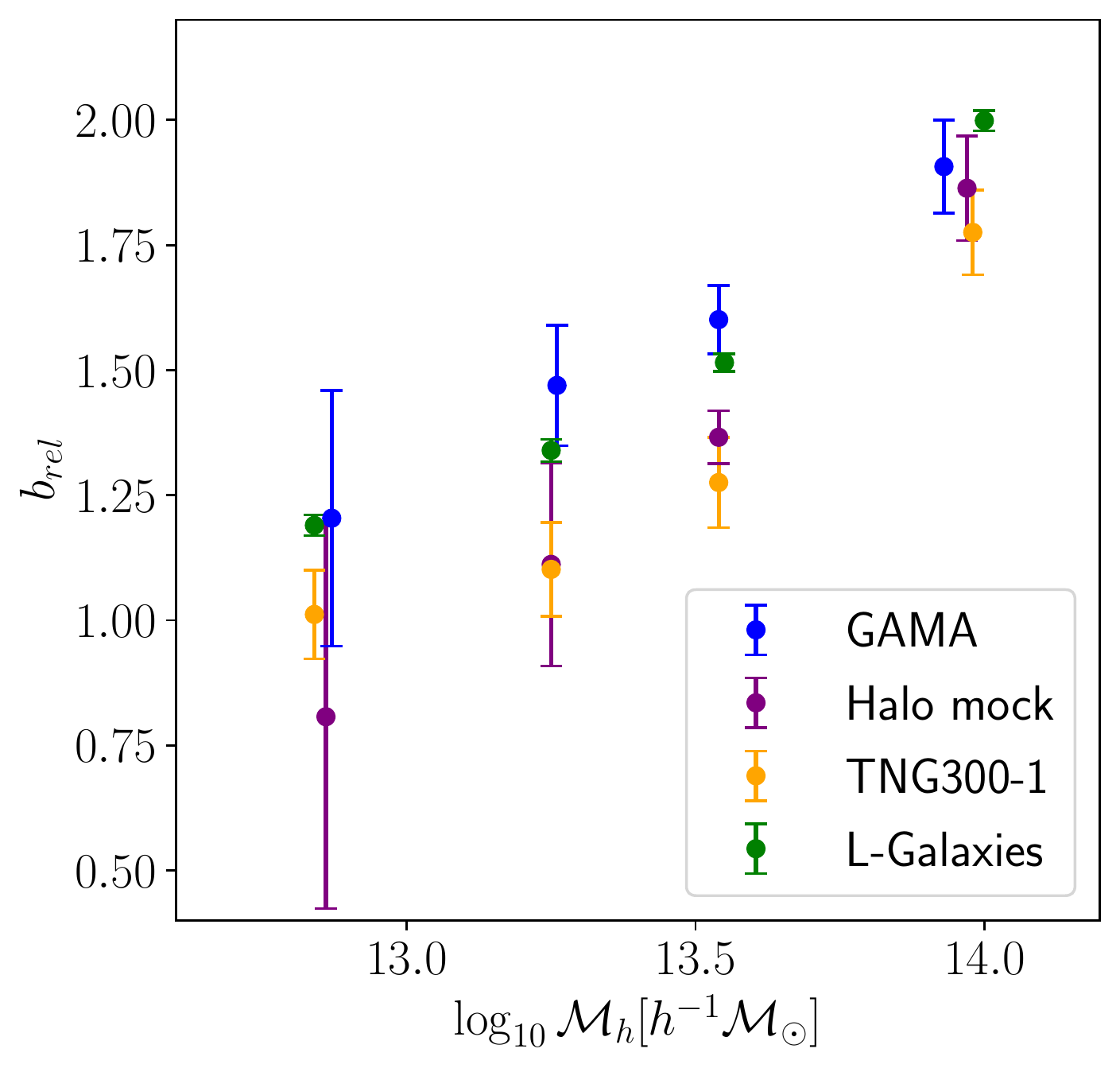}
\caption{Large-scale relative bias, averaged over scales 2--10 \hMpc,
    as a function of halo mass for GAMA, the halo mocks, TNG300-1 and \lgal.
    Uncertainties are calculated by jackknife of the average $b_{\rm rel}$ for GAMA, TNG and \lgal,
    and by scatter between the 9 realisations for the halo mocks.
}
\label{fig:mass_bias}
\end{figure}

Our finding of an increase in clustering amplitude with group mass on scales of a few \hMpc\ 
agrees with the results from the analysis of SDSS data by \cite{Wang2008}.
These authors found that the bias relative to the lowest mass bin increases quadratically with mass,
and we show a similar rise in our relative bias in Fig.~\ref{fig:mass_bias}, 
with bias averaged over scales 2--10 \hMpc. 
This trend is consistent with the results from the simulations, 
albeit with a slightly higher normalisation.
However, due to our use of different, narrower, mass bins than \cite{Wang2008},
the uncertainties from GAMA are large, and the bias values are 
not directly comparable.
In addition to the large-scale bias, we show on smaller, intra-group, 
scales, which were not considered by \cite{Wang2008},
that the dependence of clustering amplitude on group mass become significantly stronger.

This sharp increase in cross-correlation amplitude within the typical group radius
matches the results of \cite{Berlind2006},
as does evidence for a flattening of the cross-correlation on scales $r_\bot \la 0.3 \hMpc$
in our GAMA and TNG results.
\cite{Berlind2006} attribute this to either a core to the radial profile of satellite galaxies,
or to mis-identification of the centre.
We do not find evidence that the central galaxies are incorrect in our data,
so support the explanation of a central core to groups.

The result from the marked correlation functions that massive galaxies are associated with massive groups
is not surprising, and consistent with GAMA results from VM20.
More interesting is the lack of dependence of the mark on galaxy mass alone
within the radii of the smallest groups, 
in agreement with the results of \cite{Kafle2016}
that there is no mass segregation within GAMA groups.
This is in contrast to the results from SDSS, most recently in \cite{Roberts2015},
that more massive satellites are generally closer to the group centre.
Our approach of using the marked correlation is a new method to test for mass segregation,
but as our galaxy sample is volume limited in \textit{r}-band luminosity
and not in mass, our results are not directly comparable to these previous studies,
and the marked correlations must be interpreted with caution given that stellar masses 
increase with group mass.

The lack of mass segregation also suggests a breakdown in self-similarity on group scales,
as the most massive groups are found to be the most clustered, but this 
trend does not continue to galaxies within the groups.
This suggests that while on inter-group scales the galaxy distribution
depends primarily on the dark matter distribution,
within groups, baryon astrophysics has a significant effect.

\section{Conclusions}
\label{sec:close}
In this work we present group--galaxy cross-correlation
functions and mass-weighted marked correlations for the GAMA survey,
GAMA mocks, the TNG300-1 simulation, and the \lgal\ semi-analytic model.
We use four group mass bins with $12.0 < \lg {\cal M}_{h} < 14.8$
and cross-correlate with a volume-limited galaxy sample with density $5.38 \times 10^{-3} \denunit$.

We find that the group--galaxy cross-correlation function (Fig.~\ref{fig:w_p_bias})
increases systematically with group mass and with decreasing scale below $r_\bot \approx 1 \hMpc$.
There is no scale dependence on scales $r_\bot \ga 1 \hMpc$,
but the correlation amplitude still increases with group mass,
indicating that more massive groups are embedded within extended overdense structures.

Using marked correlations (Fig.~\ref{fig:marked_corr}), 
we see that the cross-correlation has the 
strongest group mass dependence at scales $r_\bot \approx 0.5 \hMpc$,
the typical group radius (defined as projected separation to the most distant
member galaxy from the group centre).
No direct dependence on galaxy mass is observed, but the combination of 
group and galaxy mass causes an enhancement over the use of group mass only.
This leads us to conclude that massive satellite galaxies are generally found in massive groups,
but do not preferentially lie close to the central galaxy.
Note that the central galaxy coincides with the iterative group centre,
and so central--group pairs are not included in the group--galaxy 
cross-correlation functions presented.

\subsection{Comparison to mocks and simulations}

We use the GAMA mock catalogues to explore the effects of systematics in the data, 
particularly the group mass estimates,
and to examine the model used for the mocks.
Comparison of mocks using friends-of-friends and halo based group
finding methods suggests that the masses may be overestimated at high redshift
and underestimated at low redshift,
although this only causes differences in the cross-correlation function
in our lowest mass bin, \mass1.

We have also compared our results against the TNG300-1 box from the 
IllustrisTNG hydrodynamical simulation and to the \lgal\ semi-analytic model.
In order to provide a fair comparison, we selected groups using a 
simple model of the GAMA selection function.

The IllustrisTNG hydrodynamical simulation agrees well with
our GAMA results in all cross-correlation bins except the lowest mass bin where
the GAMA results are least reliable.
It also displays very similar marked cross-correlations to GAMA, 
evidencing accuracy in the distribution of galaxies around groups.
The only significant difference between TNG and GAMA we see is in the 
marked galaxy auto-correlation, where the enhancement in TNG appears 
to be the same over-dependence of central galaxy mass on group mass seen in VM20.

The \lgal\ model is found to over-predict
the mass dependence of the cross-correlation,
showing an increasing bias down to the smallest scales considered.
This is seen in the marked correlations to be driven by 
stronger clustering than GAMA of the most massive galaxies,
perhaps driven by inaccurate supernova feedback.
Together with the difficulties of modelling the infall of satellites without surviving subhaloes,
this results in too many galaxies in the inner parts of the haloes.
Away from the group centre, \lgal\ shows similar group bias to GAMA,
demonstrating that the distribution of galaxies in the outer regions of 
the haloes is realistic.

\subsection{Future prospects}

While the GAMA groups are expected to be more reliable than the SDSS groups
used in previous works, due to high spectroscopic completeness and
the use of only the most reliable groups with $N_{\rm FoF} \geqslant 5$,
we are limited by the smaller area of the GAMA survey.
In future, the Wide Area VISTA Extragalactic Survey \citep{Driver2019}
is expected to be able to produce a much larger sample of galaxy groups
and so improve upon our results by reducing the uncertainties and 
allowing the use of finer mass bins.

\section*{Acknowledgements}

SDR is supported by a Science and Technology Facilities Council (STFC) studentship.
RWYMB was supported by a STFC studentship.
JL acknowledges support from the STFC (grant number ST/I000976/1).

We thank Michael Boylan-Kolchin for providing the Millennium and Millennium-II simulation
dark matter correlation functions in machine-readable form.

GAMA is a joint European-Australasian project based around a
spectroscopic campaign using the Anglo-Australian Telescope. The
GAMA input catalogue is based on data taken from the Sloan Digital
Sky Survey and the UKIRT Infrared Deep Sky Survey. Complementary
imaging of the GAMA regions is being obtained by
a number of independent survey programs including GALEX MIS,
VST KiDS, VISTA VIKING, WISE, Herschel-ATLAS, GMRT and
ASKAP providing UV to radio coverage. GAMA is funded by the
STFC (UK), the ARC (Australia), the AAO, and the participating
institutions. The GAMA website is http://www.gama-survey.org/.

We acknowledge the IllustrisTNG team for making their simulation data available.
This work used the 2015 public version of the Munich model of galaxy formation and evolution: L-Galaxies. 
The source code and a full description of the model are available at https://lgalaxiespublicrelease.github.io/

Finally, we thank the anonymous referee for providing helpful comments about the manuscript.

\section*{Data availability}

The data underlying this article will be shared on 
reasonable request to the corresponding author. 
Tabulated clustering results will be made available 
via the GAMA website http://www.gama-survey.org/.

\bibliographystyle{mnras}
\bibliography{refs.bib}

\appendix
\section{Galaxy sample statistics}
\label{appen:lf}

We desire our volume-limited GAMA, mock, TNG, and \lgal\ galaxy samples 
to have comparable clustering statistics.
In order to achieve this, they were defined to have similar number-densities 
(Table~\ref{tab:gal_def}).
Here we show the stellar mass distributions and auto-correlation functions 
of these samples.

\begin{figure} 
\includegraphics[width=\linewidth]{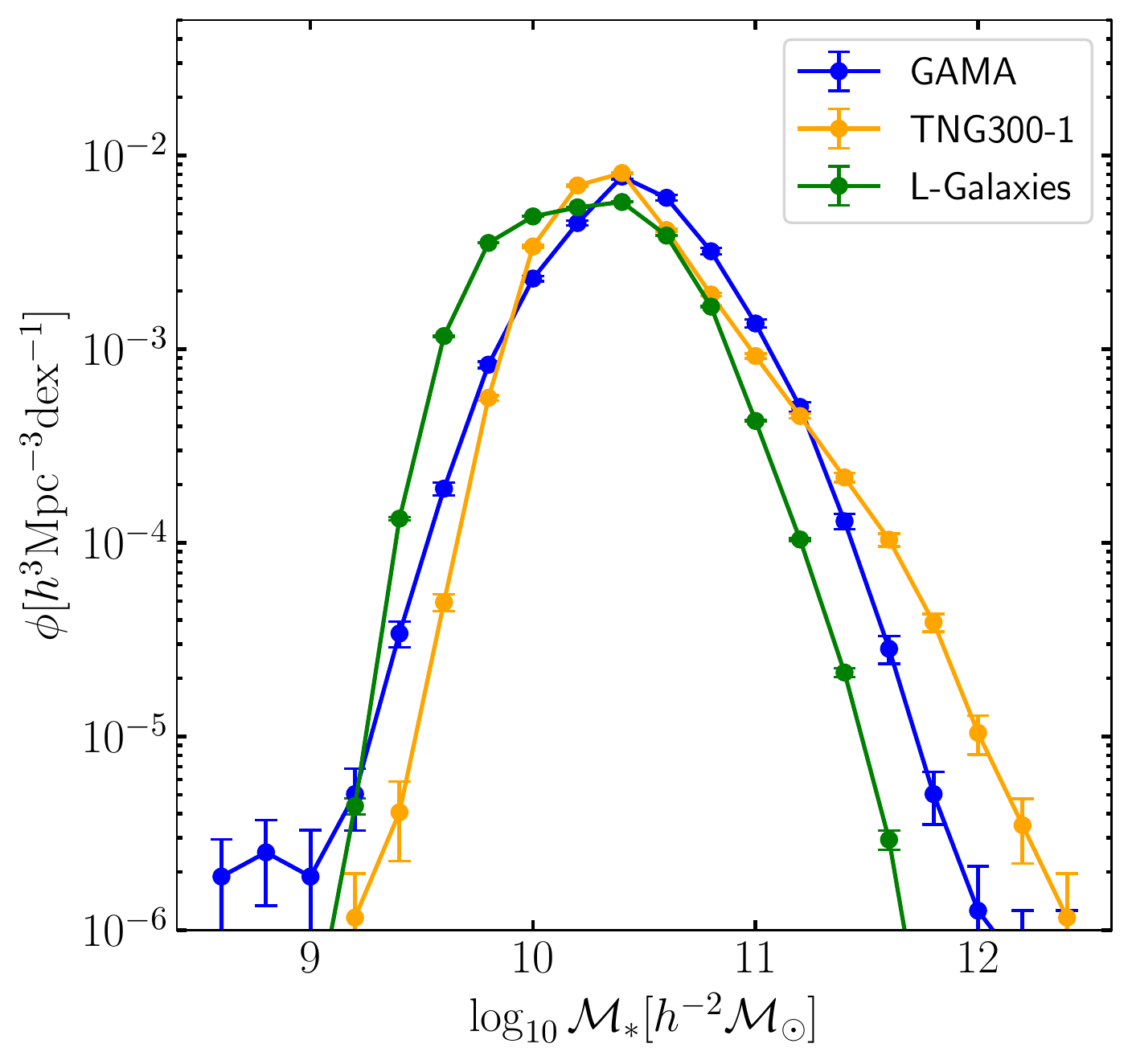} 
\caption{
  Distribution of stellar masses in our galaxy samples from GAMA, TNG and \lgal.
  The mock catalogues do not include stellar masses so are not shown.
  }
\label{fig:smf}
\end{figure}

The distributions of stellar masses in each sample (Fig.~\ref{fig:smf}) 
show some variation.
This is not surprising, as the samples are volume-limited in 
$r$-band luminosity and not in mass,
and so variations in mass-to-light ratio will affect mass-completeness.
Compared to the GAMA sample, TNG shows a narrower peak
but an over-abundance of the most massive galaxies with
$\log_{10} \mathcal{M}_* \ga 11.2 h^{-2} \mathcal{M}_{\odot}$.
\lgal\ shows a shift to slightly smaller masses than GAMA.

\begin{figure} 
\includegraphics[width=\linewidth]{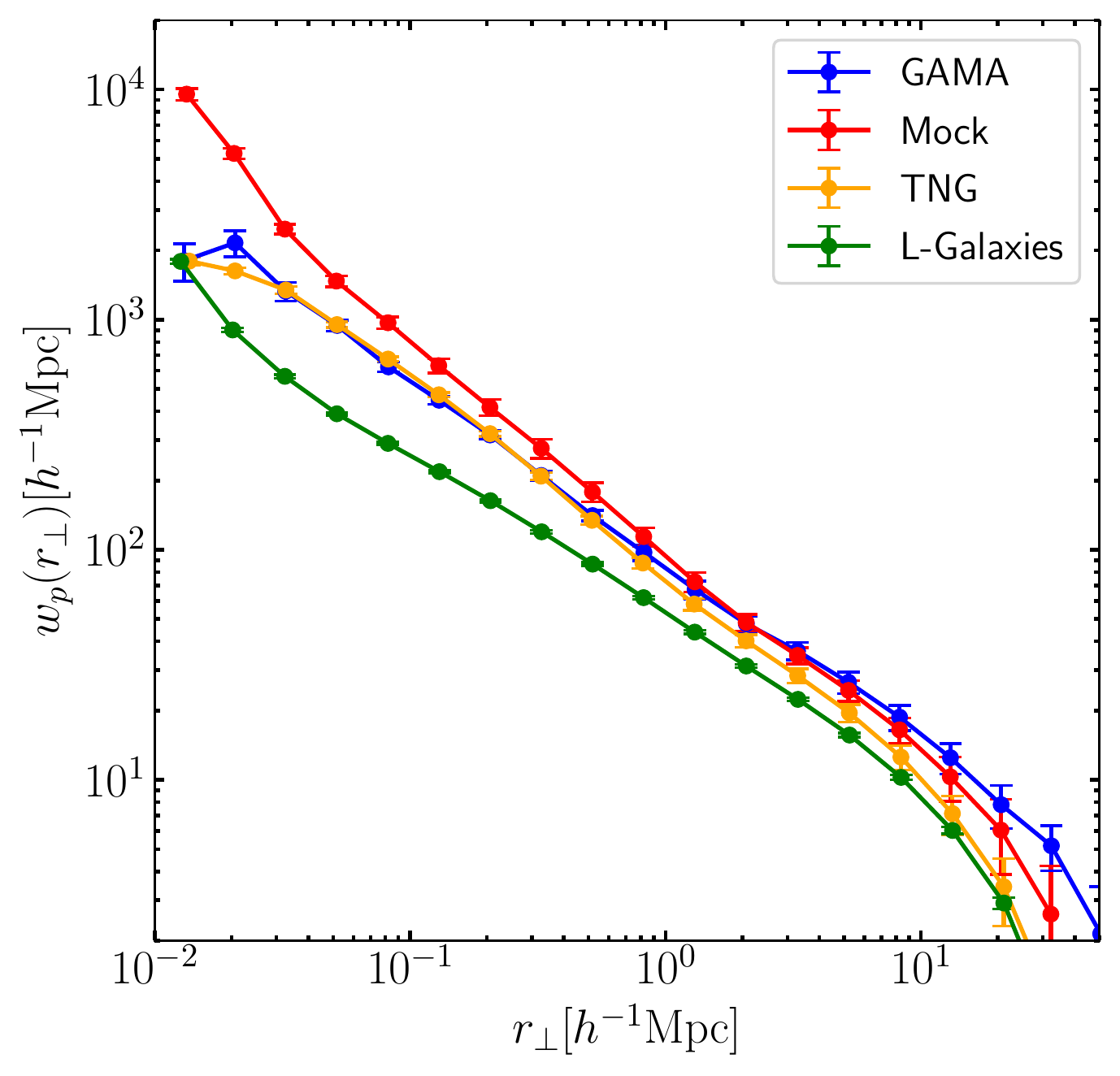} 
\caption{
  Projected auto-correlation functions of our galaxy samples 
  from GAMA, the mock catalogues, TNG and \lgal.
  }
\label{fig:autocorr}
\end{figure}

Fig.~\ref{fig:autocorr} shows the projected auto-correlation functions 
of the galaxy samples.
On small scales, GAMA and TNG agree well 
but the mocks show a slightly greater auto-correlation
and \lgal\ shows a lower auto-correlation.
On the largest scales GAMA shows the greatest clustering,
but consistent within uncertainties with the mocks.

\section{Group selection in simulations}
\label{appen:selection}

Here we compare four methods of selecting groups in mass bins
from the TNG and \lgal\ simulations,
and the effect these methods have on estimated relative bias.
The four group selection methods compared are:
\begin{enumerate}
    \item Random sampling to mimic GAMA group selection, 
    the method described in \ref{sec:sims} and used elsewhere.
    \item Spatial sampling to mimic GAMA group selection.
    Here we select groups within a distance from the origin corresponding
    to the comoving distance at which 
    the fifth brightest member galaxy would have an apparent magnitude of $m_r = 19.8$.
    This removes the periodicity of the box, and we therefore calculate the correlation function using
    the full \cite{Landy1993} estimator with random galaxies distributed around the box.
    Uncertainties on this sample are estimated using jackknife between 27 samples of equal volume
    selected by angle, and are larger than those of the random selection 
    due to the loss of periodicity.
    \item Use only of GAMA mass bin limits, without further selection.
    This results in an over-abundance of low-mass groups in a volume-limited simulation cube compared to GAMA.
    \item Adjustment of mass limits to match the mean group masses in GAMA 
    (the method employed in VM20).
\end{enumerate}

\begin{figure} 
\includegraphics[width=\linewidth]{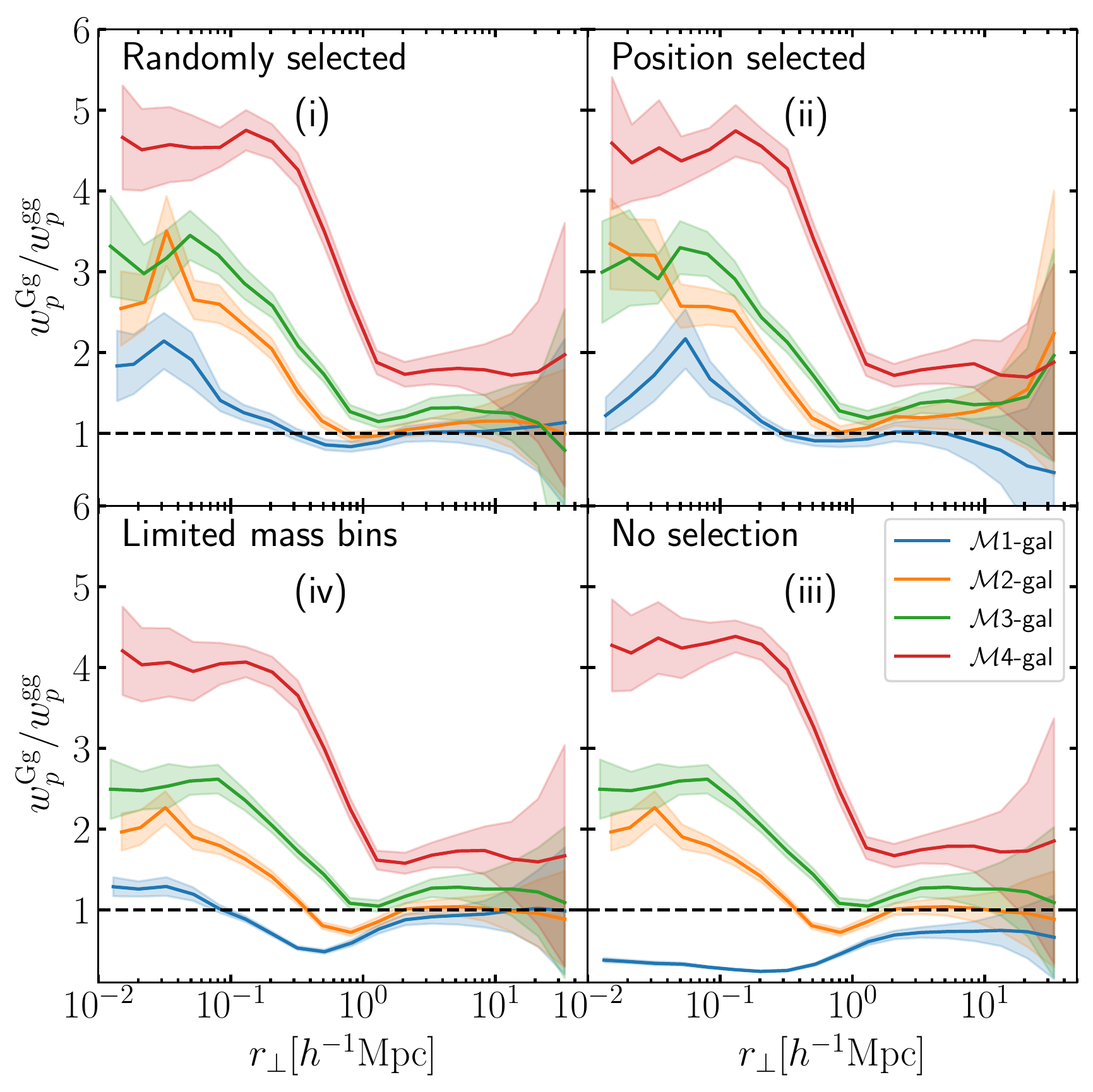} 
\caption{
  Relative bias for the 4 mass bins in the TNG simulations using 
  different selection options for groups.
  Clockwise from top left, the panels show the four group selections (i)-(iv):
  upper left, the selection of groups throughout the volume based on galaxy luminosities used in this work;
  upper right, a group selection based on galaxy luminosities and radial distance from box origin;
  lower left, the full group sample with low and high mass groups removed to match GAMA mean masses;
  and lower right, the full group sample in the volume-limited simulation.
  }
\label{fig:selection_variants}
\end{figure}

Comparing these different selection methods applied to TNG
in Fig.~\ref{fig:selection_variants},
the relative bias is consistent between the samples selected using methods (i) and (ii),
except for the smallest scales in \mass1.
Bearing in mind that the groups in sample (i) are randomly distributed throughout the
TNG data cube, whereas those in sample (ii) lie predominantly closer to the origin,
this comparison illustrates that the spatial selection of the groups has only minimal
effect on the group--galaxy cross-correlation function, and justifies our choice of
random sampling (method i).
The differences in very small-scale clustering in \mass1 likely arise from
sampling fluctuations, since the sample (ii) TNG \mass1 groups are
only taken from approximately 10\% of the total volume.

Sample (iii), lower-right panel, shows very different results.
The addition of many low-mass groups forces the bias for the lower mass bins down,
leading to anti-bias on all scales for \mass1 and near the group edge for \mass2.
This is likely due to the \mass1 TNG central galaxies having a mean luminosity $\approx 0.2$ mag
lower than the comparison galaxy sample.
Using sample (iv), lower-left panel, increases the bias for \mass1 but it still remains
below that of sample (i).

The comparison of these selection methods has demonstrated the importance of mimicking
the selection function in GAMA and validated our approach to doing so.

\section{Effect of group selection on the cross-correlation}

\begin{figure} 
\includegraphics[width=\linewidth]{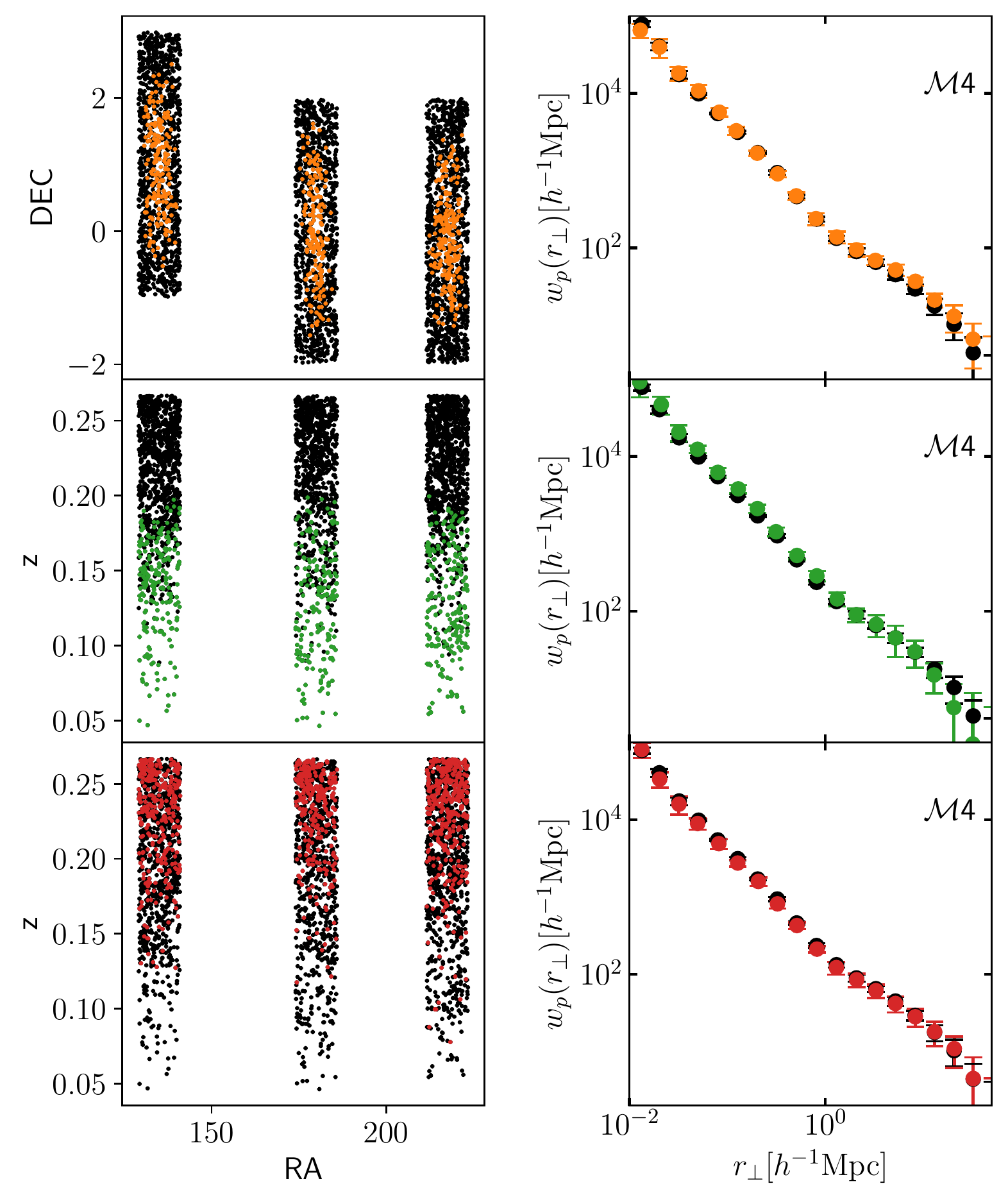} 
\caption{
  The effect of group selection on the group--galaxy cross-correlation function 
  in the FoF mock catalogue for bin \mass4.
  Left panels show the selected groups and right panels show the resulting 
  cross-correlation, with black points in all cases showing the full sample.
  The upper row shows a selection excluding groups near the field edges,
  the middle panels show low-redshift groups
  and the lower panels high-redshift groups.
  }
\label{fig:x_corr_test}
\end{figure}

We show here that our GAMA cross-correlation results are not significantly affected
by group selection effects.
Fig.~\ref{fig:x_corr_test} shows the cross-correlation for the \mass4 bin
in the FoF mock with different artificial selection effects introduced.

To check the effects of missing groups near the field edges,
we select groups based on the distance from the field centres.
This results in a slight increase in cross-correlation amplitude
on large scales, but consistent within uncertainties.
We also show the effects of selecting low- and high-redshift groups.
There are no significant shifts in either case.

The similarity of all the cross-correlations shown here 
(and similar results are obtained for the other mass bins and the halo mock)
demonstrates that our results are robust to the effects of group selection.

\section{Marked correlations by rank}
\label{appen:ranks}

\begin{figure*} 
\includegraphics[width=\linewidth]{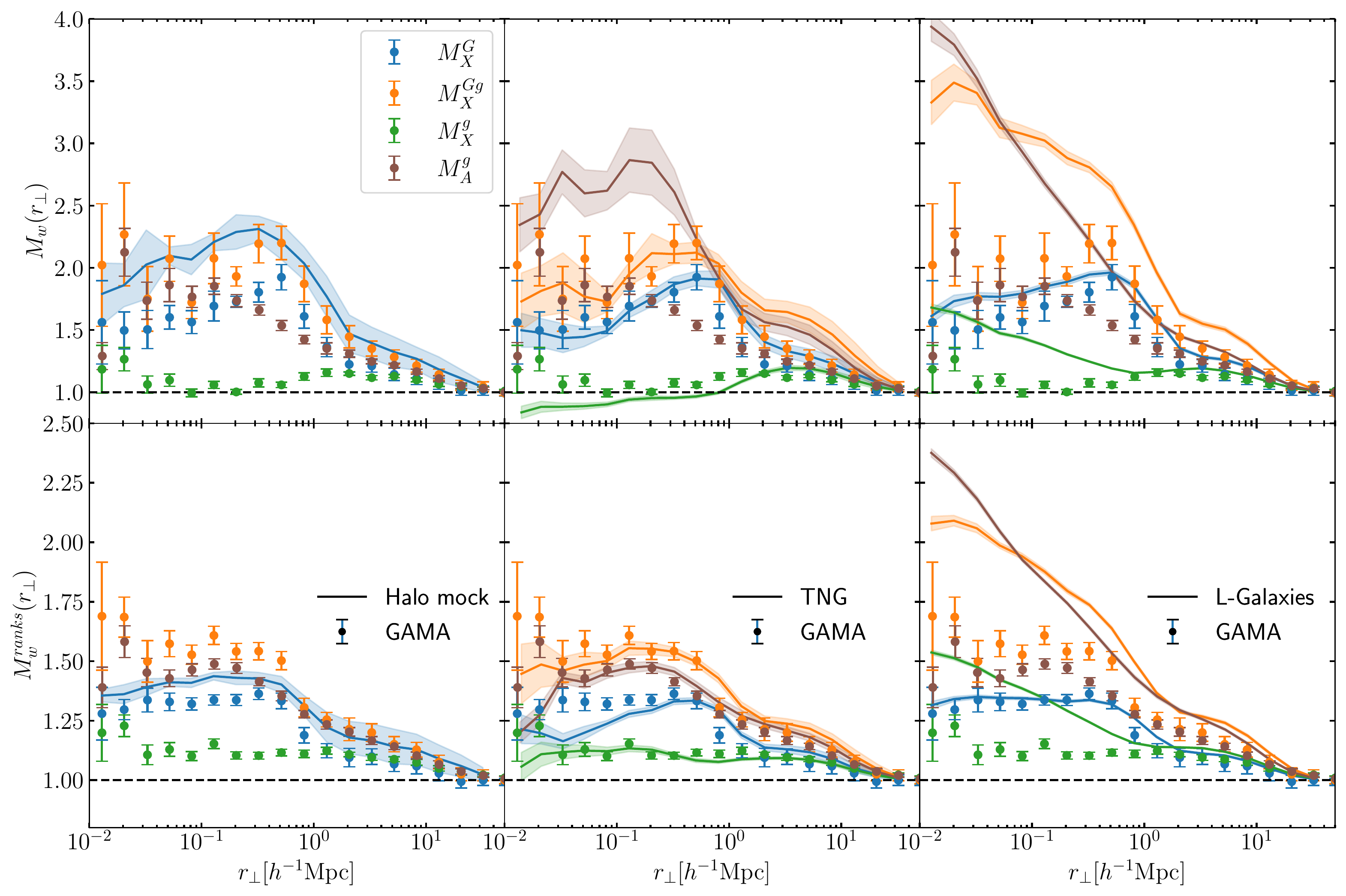} 
\caption{
  Marked correlations using masses and the rank ordering of mass.
  Upper panels are the same as the lower panels of Fig.~\ref{fig:marked_corr},
  lower panels show the results from rank ordering the masses.
  Symbols and error bars show the GAMA results in all panels;
  lines of corresponding colour show results from the halo mock in the left panels,
  the Illustris TNG300-1 simulation in the central panels and \lgal\ in the right panels.
  }
\label{fig:ranked_corr}
\end{figure*}

In order to check the effect of our choice of galaxy or group mass as a mark, 
we perform an alternative marking using the rank ordering method of \cite{Skibba2013}.

We sort the masses in ascending order and assign the rank as the position
in the sorted list.
Results from using these ranks as marks are shown in Fig.~\ref{fig:ranked_corr}.
When compared to the marked correlations using masses shown in Fig.~\ref{fig:marked_corr},
it is clear that the amplitude of the marked correlations is reduced when using 
ranks. However, the qualitative comparison between different weighting options and samples
remains the same.

The most notable difference is the TNG galaxy mass-weighted auto--correlation.
In that case, using rank orderings brings the mark into agreement with GAMA on most scales,
suggesting that the enhanced mark seen in Fig.~\ref{fig:marked_corr} is due to the differences
in the shape of the stellar mass function between TNG and GAMA in Fig.~\ref{fig:smf}.

The other visible difference is that the cross-correlation weighted by galaxy masses 
is greater than 1 when using ranks for GAMA and TNG.
However, there is no scale dependence, meaning this is not a signal of mass segregation.
Instead it appears to confirm the galaxies from our sample which are in the groups
have slightly higher masses than the average of the volume limited sample.

\bsp

\label{lastpage}

\end{document}